**Autonomous LLM-driven research from data to human-verifiable research papers**

Tal Ifargan[1,*], Lukas Hafner[2,*], Maor Kern[3], Ori Alcalay[3] and Roy Kishony[2,4,5]

[1] Faculty of Data and Decision Sciences, Technion–Israel Institute of Technology, Haifa, Israel.

[2] Faculty of Biology, Technion–Israel Institute of Technology, Haifa, Israel.

[3] Epsio, Tel Aviv, Israel.

[4] Faculty of Computer Science, Technion–Israel Institute of Technology, Haifa, Israel.

[5] Faculty of Biomedical Engineering, Technion–Israel Institute of Technology, Haifa, Israel.

# Abstract

As AI promises to accelerate scientific discovery, it remains unclear whether fully AI-driven research is possible and whether it can adhere to key scientific values, such as transparency, traceability and verifiability. Mimicking human scientific practices, we built data-to-paper, an automation platform that guides interacting LLM agents through a complete stepwise research process, while programmatically back-tracing information flow and allowing human oversight and interactions. In autopilot mode, provided with annotated data alone, data-to-paper raised hypotheses, designed research plans, wrote and debugged analysis codes, generated and interpreted results, and created complete and information-traceable research papers. Even though research novelty was relatively limited, the process demonstrated autonomous generation of *de novo* quantitative insights from data. For simple research goals, a fully-autonomous cycle can create manuscripts which recapitulate peer-reviewed publications without major errors in about 80-90%, yet as goal complexity increases, human co-piloting becomes critical for assuring accuracy. Beyond the process itself, created manuscripts too are inherently verifiable, as information-tracing allows to programmatically chain results, methods and data. Our work thereby demonstrates a potential for AI-driven acceleration of scientific discovery while enhancing, rather than jeopardizing, traceability, transparency and verifiability.

# Introduction

Recent advances in natural language processing have resulted in LLMs, such as ChatGPT, capable of writing text, answering questions, and generating code at a human level (*1–5*). Furthermore, augmenting LLMs with external tools as well as automating iterative algorithmic prompting and multi-agent interactions has enabled tackling even more complex, multi-step, tasks such as solving math problems (*6–8*), coding and debugging large code projects (*9, 10*), and creating book-long texts and scripts (*11*). Most recently, LLMs have even demonstrated a capacity of designing and running experimentation as well as performing clinical diagnostics (*12–14*). Yet, despite all these advances, scientific research, and in particular the *de novo* creation of insights from data, remains a stronghold of human intelligence and ingenuity (*15–20*). The recent advancement of AI has led to a vivid discussion on the potential and risks of their application in science (*21*), and to emerging guidelines, emphasizing the importance of key values including accountability, oversight and transparency, notoriously challenging in AI (*22*).

Conducting research and compiling results and conclusions into a transparent and methodologically traceable and verifiable scientific paper is a highly challenging task, involving multiple interconnected steps and requiring planning, inference, and deduction, as well as the meticulous tracing of information. While scientists may in principle follow an infinite number of creative paths towards discovery, certain conventional research paths have been established (*23*). In particular, such paths typically follow these almost-canonical steps: formulating and reshaping a research question in light of the literature, designing and executing a research plan, interpreting the results in the context of prior knowledge, and writing a research paper. Beyond its established multi-step structure, human-driven scientific process has three additional key characteristics. First, the process is not linear, it often requires iteratively setting back to earlier steps. Second, it is built on a rigorous tracing and control of both textual and quantitative information among steps. Finally, at each of the steps, human scientists rely on feedback from peers, mentors, or external reviewers, enabling an overall collective strength beyond individual capabilities. Taken together, these key features make science a unique process of human creativity.

Here, inspired by how research is conducted by human scientists, we build data-to-paper, an automation platform that systematically guides multiple LLM and rule-based algorithmic agents through the conventional steps of scientific research, with automated feedback, iterative cycles of review and revision, and with structured control and tracing of information flow among these research steps. We specifically focus on a relatively simple and well-defined process of hypothesis-testing research. Starting with a human-provided dataset, the process is designed to raise hypotheses, write, debug and execute code to analyze the data and perform statistical tests, interpret the results and write well-structured scientific papers which not only describe results and conclusions but also transparently delineate the research methodologies, allowing human scientists to understand, repeat and verify the analysis. The discussion on emerging guidelines for AI-driven science (*22*) have served as a design framework for data-to-paper, yielding a fully transparent, traceable and verifiable workflow, and algorithmic "chaining" of data, methodology and result allowing to trace downstream results back to the part of code which generated them. The system can run with or without a predefined research goal (fixed/open-goal modalities) and with or without human interactions and feedback (copilot/autopilot modes). We performed two open-goal and two fixed-goal case studies on different public datasets (*24–27*) and evaluated the AI-driven research process as well as the novelty and accuracy of created scientific papers. We show that, running fully autonomously (autopilot), data-to-paper can perform complete and correct run cycles for simple goals, while for complex goals, human co-piloting becomes critical.

## Implementation

To autonomously analyze a provided dataset and create a research paper, data-to-paper guides multiple LLM and rule-based agents through a series of pre-defined "research steps", each designed to create well-defined quantitative or textual "research products" (Fig. 1). The process includes the following steps: data exploration, literature search and iterative formulation of a research goal and hypothesis, creating a hypothesis testing plan, writing data analysis code, creating scientific tables, searching related literature, and writing the paper section by section (Fig. 1A; Fig. 1B, top; in total 17 steps). The research goal can also be provided as human input, in which case the goal-determining steps are skipped ("Fixed-goal modality", dashed bypass arrow, Fig. 1A; Methods). The process runs automatically through the series of steps (with human overseeing and approval; Methods), with each step creating one or more research products, of different types ("Free text", "LaTex text", "Structured text", "Binary decision", "Citations", "Python code" and "Numerical data"; Fig. 1B, left). In coding steps, the LLM creates a Python code, which is then executed by data-to-paper to analyze the provided dataset and create numerical data products (like tables for the paper; Methods). In literature-search steps, a structured list of queries created by the LLM is used to retrieve a list of citations from an external citation database (*28*) (Fig. 1A; Methods). Ultimately, these intermediate products are automatically assembled into a complete research paper (labeled with an AI-created watermark for transparency; Fig. 1B; Methods).

Each research step is implemented as a distinct conversation, with agent identity specification, provision of prior research products, mission instructions, and LLM responses with iterative feedback (Fig. 1C; Methods; Fig. S1). First, the LLM agent is designated a specific identity (Methods; e.g. "*You are a scientist who needs to write literature search queries*"; "Performer system prompt", Table S1). Next, data-to-paper populates the conversation with a set of "provided prior products": a list of messages providing the LLM with a pre-defined subset of research products of prior steps, deemed important for the focal task (Fig. 1B; Fig. S1; "Provided prior products", Table S1). This rigor control of information flow among steps minimizes possible hallucinations due to mixing relevant and irrelevant information (*29*). It also allows data-to-paper to trace, verify and chain the sources of numeric results cited by the LLM (Methods). Next, a step-specific "mission prompt" message is appended, defining the new product that the LLM is expected to create (e.g. "*Please write literature-search queries...*"; "Performer mission prompt", Table S1). Then, data-to-paper requests a response from the LLM model API (*30–32*), from which it extracts the requested product (based on defined formatting; Fig. S1; Table S1; Methods). The extracted product then undergoes a series of rule-based algorithmic checks, providing constructive feedback to the LLM upon failure (Methods; Fig. 1C; Fig. S1). In particular, to minimize errors in the coding steps, we have built a unique framework that imposes guardrails against commonly observed coding and statistics analysis errors, through a series of static code checks, runtime errors, package-specific guardrails, and output verifications (Fig. 1A, "Coding" block; Methods).

Once the created product passes rule-based review, it may further be refined through LLM review (*9*, *33–38*) (steps with "Review" ellipses, Fig. 1A; Methods). LLM review is implemented as a parallel, role-inverted conversation, effectively creating an exchange between two LLM agents (Methods; Fig. 1C; Fig. 2A; Fig. S2). In co-pilot mode, the human user can provide additional review comments, resulting in further LLM iterations (Methods). Once a product passes rule-based, LLM and optionally a human review, the step is concluded and data-to-paper proceeds to the next step, until all products are created and the paper is assembled. While data-to-paper can work with any LLM, in practice, our implementation uses ChatGPT; using the current state-of-the-art open-source LLMs leads to frequent mistakes that preclude completing full research cycles (Methods; Table S2; Fig. S3). Of note, since ChatGPT is not a deterministic model, each run of data-to-paper, even on the same dataset and either with or without a human-provided goal, unfolds with different analyses, yielding different overall manuscripts.

## Open-goal research on public datasets

Running in an open-goal, autopilot modality, we provided data-to-paper with two publicly available datasets: (*i*) "Health Indicators" dataset (*24*), an unweighted curated subset of the CDC's Behavioral Risk Factor Surveillance System (BRFSS) from 2015 (*39*), with 253,680 clean responses, each including 22 features related to diabetes and general health, and (*ii*) "Social Network" dataset (*25*), a directed graph representing Twitter interactions among members of the 117th US congress, as well as member affiliations (Chamber, Party and State). For each of these two datasets, we ran data-to-paper for 5 full research cycles, creating 10 distinct manuscripts (Supplementary Dataset A,B; Supplementary Data Descriptions A,B; Supplementary Manuscripts A1-5, B1-5). During these research cycles, which took about an hour each, data-to-paper generated and corrected hypotheses, created and debugged code, composed search queries and retrieved citations, and wrote and revised the manuscript section by section (full conversations in Supplementary Runs A1-5, B1-5; Fig. 2B; Figs. S4-S7). All created manuscripts properly followed the canonical structure of a research paper, including a proper title and abstract, a well-formulated introduction that stresses the research questions in light of relevant literature, a method section providing a transparent and human-traceable description of the analysis and key methodologies, several supplementary sections providing all custom-written codes, properly formatted scientific tables, a results section which describes the findings while properly referring to each of the tables, and a referenced discussion section which summarizes the results, delineate limitations and puts the findings in a broader context (Supplementary Manuscripts A1-5, B1-5). While similar in structure, the 5 different papers produced for each dataset addressed different topics and raised and tested different hypotheses (Table 1, Table S3). These papers are not highly creative, yet they do define a reasonable set of hypotheses, test them with simple straightforward statistical approaches, and ultimately create and adequately report *de novo* insights from the provided data.

Manually vetting the data analysis and the text of these papers, we found that out of these 10 open-goal papers, 8 reported correct analysis with only minor wording imperfections, yet 2 were erroneous, showing fundamental analysis or interpretation mistakes (Supplementary Manuscripts A1-5, B1-5). The analyses in all 5 "Health Indicators" papers were based on either logistic or linear regression models, all adequately performed while accounting for a reasonable choice of confounding factors (Table S4). Furthermore, interaction terms were adequately added when needed, and the dataset was adequately restricted to reflect the tested hypotheses (restricting to the diabetic sub-population; Table S4). For the "Social Network" dataset, papers were based on linking graph properties with node properties, as well as on creating new node properties (e.g. State representation size), and then applying linear regression, ANOVA, or Chi-square on either the graph nodes or edges as appropriate (Table S4; see methods sections and analysis codes in each of the created papers, Supplementary Manuscripts A1-5, B1-5). In all 10 papers, the generated scientific tables correctly represented the results of the analysis. Vetting the text, we observed that data-to-paper is adequately interpreting the analysis results with factual statements, correctly referring to tables and citing key numeric values from the analysis, and reasonably describing the research question and findings in the context of existing literature (green highlights, Supplementary Manuscripts A1-5, B1-5; Methods). We also detected multiple imperfections, such as generic phrasing, overstatement of novelty, and inadequate and sometimes lacking choice of citations (yellow and orange highlights, Supplementary Manuscripts A1-5, B1-5). More major, result-affecting, mistakes were found in 2 of the 10 papers: In one of the "Health Indicators" papers, a correct analysis was misinterpreted due to hallucinations in the goal specification step, leading to conclusions beyond the scope of the analysis; and in one of the "Social Network" papers, an erroneous analysis was performed, resulting in unfounded statements on statistical associations between social interactions and party affiliations (red highlights, Supplementary Manuscript A2 and B2, respectively).

## Estimating reliability in reproducing peer-reviewed results

To more systematically assess its error rate in autopilot mode, we applied data-to-paper in a fixed-goal modality in two case studies for which we have benchmarks of published peer-reviewed results. We specifically wanted to check two critical aspects for the reliability of analysis and interpretation: the proper reporting of both positive and negative findings (challenge 1), and the performance for tasks with multiple different steps with tunable breadth (challenge 2). To test data-to-paper capacity in these two challenges, we chose the following two examples of peer-reviewed studies: a study by Saint-Fleur et al. (*26*), which adequately reports both positive and negative findings related to the association of a policy change in a Neonatal Intensive Care Unit with treatment choice and treatment outcome, respectively (challenge 1); and a study by Shim et al. (*27*), which builds several Machine Learning models for predicting optimal intubation depth in pediatric patients, and compare their prediction accuracy with formula-based models, thereby requiring multiple analysis steps, whose breadths can be gradually tuned (by altering the number of models to compare; challenge 2). Both studies provide well-annotated datasets and both were published after the knowledge cutoff date of the ChatGPT models that we used (September 2021; Methods). For each of the two case studies, we have provided data-to-paper with the research goal of the original publication and the corresponding dataset and ran it for 10 independent research cycles (Supplementary Data Descriptions C,D; Supplementary Datasets C,D; Tables S5,S6; Supplementary Manuscripts C1-10, Da1-10). Within each case study, the created papers were all similar to each other in their content, terminology and structure. Indeed, quantifying content similarity by the pairwise cosine distance between the vector embeddings of the title and abstract of all created manuscripts (*40*) showed tight and distinct clusters corresponding to the 4 case studies (two open-goal and two fixed-goal; Fig. 3). Furthermore, the fixed-goal papers were also similar to their respective original studies (*26, 27*) in content, terminology and in their vector embeddings (Fig. 3).

We manually vetted the analysis and reported results of the manuscripts created for each of the two study-reproducing challenges. For challenge 1, we found that all papers correctly reproduced the analysis, and 8 of them reached the overall correct conclusions and adequately reported both the negative and positive results. All of these manuscripts used adequate statistical methodologies, either matching the methods used in the original study (*26*) or providing valid alternatives (Table S5; Supplementary Manuscripts C1-10, Supplementary Runs C1-10). Yet, despite correct analysis, in 2 out of these 10 papers we identified interpretation errors, which in one of the papers also affected the overall conclusions (Fig. 4; Supplementary Manuscripts C1,2, red and orange highlights; Tables S5,S6). In challenge 2, we found that the rate of error critically varied with the breadth of the analysis; while data-to-paper frequently failed when presented with the original, broad research goal (90% error rate), it was able to correctly perform this multi-step model development research for almost identical research goals except for requesting fewer models (10-20% error rate; Fig. 4). We note that as the breadth of the task increases, the number of iterations required to complete the Data Analysis step increases, providing a potential possibility to alert of too complex analysis and difficult goals (Fig. 4, bottom). We further note that in all cases, the process reliability depends on the formulation of the research goal and the description of the dataset; less detailed and explicit formulations can increase analysis errors (Fig. S8; Supplementary Manuscripts Dai1-10, Dbi1-10, Dci1-10 and Data Descriptions Dai, Dbi, Dci). Finally, allowing human co-piloting (Methods), a 2-3 single-sentence review comments per run, typically in the code writing step, allows creating accurate papers consistently even for the more complex goals (Fig. 4; Supplementary Human Co-piloted Manuscripts 1-3). Altogether, these case studies provide an assessment of data-to-paper's analysis and interpretation reliability, showing that for simple research goals it can autonomously create reliable manuscripts in 80-90% of the cases, and that for more complex goals human-copiloting is critical to assure reliability.

Finally, noting the effort and necessity of manually vetting and verifying created manuscripts, we harnessed data-to-paper step-to-step information tracing to chain results, methodology and data in created manuscripts through algorithmically verified hyperlinks (Methods). This approach creates manuscripts in which all cited numeric values are recursively linked to the specific lines of code where they are created. In particular, numbers cited in the manuscript are linked to a "Notes" appendix providing their formula and its explanation, and from there to the specific table where values used in these formula have originated from, and from the table to the corresponding output file of the code from which the table was created, and from there to the very specific part of code which produced this output file (see clickable hyperlinks in Supplementary Data-chained Manuscripts A-D). Such data-chained manuscripts facilitate systematic vetting of papers, setting a new standard for traceability for the coming era of AI-powered research.

## Discussion

Inspired by key features of human research, we use prompt automation, tool augmentation, and multi-agent interaction approaches (*9*, *12*, *33*) to guide multiple LLM agents through a full research path leading from annotated data to well-structured transparent, human-verifiable papers. Tracing information through the different research steps allows data-to-paper to create "data-chained" manuscripts, where results, methodologies and data are programmatically linked. While the novelty of this AI-driven research falls well behind high-end contemporary science, it did demonstrate a *de novo* creation of new insights from provided data, thereby recapitulating a key aspect of human research and taking science automation well beyond what is possible with algorithmic data exploration (*41*). Furthermore, the process demonstrates versatility with respect to data types and research domains, and is able to produce different forms of scientific output, such as association studies, network analysis, or development and testing of machine learning models. Run fully autonomously, the process however is not error-free; despite minimizing errors with multiple guardrails, algorithmic checks, review cycles, and tight control of information flow, the notorious problem of LLM hallucinations (*29*) leads to fundamental errors in about 10 to 20 percent of created papers, for simple analysis tasks, and to consistent failure for more complex tasks. Integrating human co-piloting, few short review comments were sufficient to overcome errors even for complex tasks. Our current implementation has multiple constraints: it is limited to textual and table outputs, is unable to formulate and pursue follow-up questions, and is limited to hypothesis-testing research.

Despite these current limitations, the ability of LLMs to carry out scientific research presents important opportunities, but also major challenges. Indeed, such AI research approaches, capable of creating *de novo* research papers from data in just an hour, could dramatically accelerate the pace of the scientific process. However, there are also risks associated with this development, such as the dishonest use of such systems, e.g. in the context of *P*-hacking (*42*), or overloading the publication system with medium-level and generic manuscripts addressing insignificant problems (*43–45*). Our approach implements specific features to mitigate some of these risks, in line with emerging guidelines on AI in science (*22*), including a transparent, overseeable process allowing human co-piloting, unbiased reporting of either positive or negative results, and the creation of transparent, AI-marked, "data-chained" and human-verifiable papers. Given the relatively limited novelty and the potential for errors in AI-driven research, as well as the need for ethical judgments and accountability (*22*), we anticipate and urge that such AI-driven approaches will used as scientist co-pilots, helping scientists in the more straightforward tasks, thereby allowing them to focus their minds and creativity on higher-level concepts.

## Methods

**Datasets.** We used 4 datasets, each consisting of data files ("Data", Fig. 1B; Supplementary Datasets A-D) and metadata items (the human-provided products "Data file description" and "General description of dataset", Fig. 1B; Supplementary Data Descriptions A-D). (A) "Health Indicators" dataset (*24*). A clean unweighted subset of CDC's Behavioral Risk Factor Surveillance System (BRFSS) 2015 annual dataset (*39*), downloaded from Kaggle (*24*). It contains 253,680 survey responses each with 22 features related to diabetes and different health indicators, with no missing values. No change in the dataset was made; data-to-paper was provided with the csv file as downloaded from Kaggle. (B) "Social Network" dataset (*25*). A directed graph of Twitter interactions among the 117th Congress members (*25*). Two data files were provided to data-to-paper: (*i*) a csv file containing a list of directed unweighted edges, representing Twitter engagements among Congress members (downloaded from Stanford Network Analysis Project (*46*), with the weights removed), and (*ii*) a csv file containing the affiliations of each Congress member, including their Chamber, Party and State (downloaded from FRAC (*47*)). (C) "Treatment policy" dataset (a test case to reproduce Saint-Fleur et al. (*26*)). A dataset on treatment and outcomes of non-vigorous infants admitted to the Neonatal Intensive Care Unit (NICU), before and after a change to treatment guidelines was implemented. As input to data-to-paper, the file downloaded from Saint-Fleur et al. (*26*) was converted into a csv file, with minor cleanups: converting column headers into alphanumeric names, converting string binary into integer binary, and removing the following irrelevant columns: 'RACE', 'RACE IN TWO CATEGORIES', 'ETHNICITY', 'Singleton /Multiple', 'Maternal Diabetes…', 'PRETERM VS TERM', 'ROUTINE RESUSCITATION…', 'Respiratory Support', 'Exposure to xrays', 'X-Ray finding' (without removing these columns, the "Data exploration" step of data-to-paper occasionally created too large output files leading to breaking the token limit of ChatGPT). (D) "Treatment Optimization" dataset (a test case to reproduce Shim et al. (27)). A dataset of 967 pediatric patients, which received mechanical ventilation after undergoing surgery, including an x-ray-based determination of the optimal tracheal tube intubation depth and a set of personalized patient attributes to be used in machine learning and formula-based models to predict this optimal depth. As input for data-to-paper, we removed irrelevant columns, leaving only the ones used in the original study: 'tube', 'sex', 'age_c', 'ht', 'wt', 'tube_depth_G'. For datasets C and D, we further provided data-to-paper with the research goal of their respective original studies. Research goals and dataset descriptions have been formulated in an iterative and empirical process: We consulted with ChatGPT on best phrasing and terminologies, tested them in pilot runs, identified misunderstandings and vague or ill-defined statements, and adapted the descriptions accordingly. Dataset descriptions and file descriptions for all datasets, as well as the research goal for datasets C,D, are provided in Supplementary Data Descriptions A-D.

**Execution of data-to-paper.** For each run, data-to-paper is provided with a dataset, its associated metadata, and an optional research goal and proceeds automatically through the stepwise research process (Fig. 1A,B). In open-goal modality, data-to-paper runs through the entire research process (Fig. 1A,B). In fixed-goal modality, the research goal is provided and the steps for choosing a research goal are skipped ("Fixed-goal modality", Fig. 1A). Human interactions are implemented as a simple user approval at each research step (autopilot mode; user is only overseeing) or with complete human review through an interactive app (co-pilot mode; user can provide reviewing comments at each step). For each dataset, we performed multiple data-to-paper runs, as follows. (A) "Health Indicators" dataset. We ran data-to-paper in an open-goal modality with this dataset and its associated metadata for 5 full research cycles (Supplementary Runs and Manuscripts A1-5). Overseeing the process, we aborted and restarted the 5th run 3 times after the "Goal validation" step, when observing that the chosen Research goal was too similar to goals of prior research cycles. (B) "Social Network" dataset. We ran data-to-paper in an open-goal modality with this dataset and its associated metadata for 5 full research cycles (Supplementary Runs and Manuscripts B1-5). To minimize overlapping goals in repeated runs, a list of the already-chosen previous goals was presented as part of the "mission prompt" of the "Research goal" step. (C) "Treatment Policy" dataset. We ran data-to-paper in a fixed-goal modality for 10 research cycles with this dataset and its associated metadata and research goal (Supplementary Runs and Manuscripts C1-10). (D) "Treatment Optimization" dataset. We ran data-to-paper in a fixed-goal modality for 10 full research cycles with this dataset and its associated metadata and research goal (Supplementary Runs and Manuscripts Da1-10). We then ran data-to-paper with 5 modified research goals (Supplementary Data Descriptions Db, Dc, Dai, Dbi, Dci) for 10 times per goal (Supplementary Runs and Manuscripts Db1-10, Dc1-10, Dai1-10, Dbi1-10, Dci1-10). As these additional 50 runs were only used to annotate analysis failure, we terminated them after the "Title & abstract" step (to save unnecessary api calls). In addition, we ran data-to-paper in co-pilot mode for three times on the original goal (Supplementary Data Description Da). During each of these runs, we provided several review comments, typically in the code writing step (Supplementary Human Co-piloted runs 1-3).

**Overview of data-to-paper implementation.** We implement data-to-paper as a chained list of research steps, each designed to create one or more research products based on a provided subset of prior research products (Fig. 1A,B). Each such research step is implemented as a distinct "Performer conversation", which specifies LLM identity, relevant prior research products and a step-specific "mission prompt" requesting the LLM to create a focal product. Product extracted from the LLM response undergoes rule-based review and programmatic feedback requesting corrections is sent back to the LLM. For certain research steps, once the product passes rule-based review it can also be sent for LLM review, which is implemented in a parallel "Reviewer conversation" ("Review", "LLM reviewer agent" in Fig. 1A,C respectively). The research step terminates with a final product that has passed both rule-based and LLM-based review. Once all steps are completed, a manuscript is assembled and compiled from the products of all relevant steps.

**Devising prompts.** The prompts used by data-to-paper in each of the research steps are listed in Table S1. These prompts have been designed in an iterative and empirical process. First, we devised an initial version for each of the prompts, focusing on the key aspect of their focal task (dark brown text, Table S1). Additionally, we added to each prompt formatting instructions for the research product (light blue and red text, Table S1). Then, we tested ChatGPT responses through multiple pilot runs, identified wrong or inadequate responses, and adapted the prompts with additional details and specifications (light brown text, Table S1). In cases where ChatGPT still failed to consistently respond as expected, we also added one-shot examples (green text, Table S1).

**Message types.** Messages in a conversation are designated as either SYSTEM, USER, or ASSISTANT (per OpenAI API terminology (*30*)). SYSTEM and USER messages are programmatically composed by data-to-paper. ASSISTANT messages are created by the LLM. We also implement LLM-surrogating ASSISTANT messages, which are messages created programmatically by data-to-paper, but are attributed to the ASSISTANT (namely, they appear to the LLM as if they were created by it).

**Performer conversation.** At the onset of each research step, a distinct Performer conversation is initiated and programmatically pre-filled with a list of "context messages": (*i*) "system prompt" defining the identity of the performer LLM agent ("Performer system prompt", Table S1); (*ii*) "provided prior products", a list of USER messages providing the LLM with a pre-defined subset of research products of prior steps, with each such USER message followed by an LLM-surrogating acknowledgment message (Fig. 1B, Figs. S1,S4; "Provided prior products", Table S1); and (*iii*) USER message describing to the LLM what it is requested to do in the current step ("Performer mission prompt", Table S1). This pre-filled Performer conversation is then sent to the LLM API (*30–32*) to request an initial response (Figs. S1,S4). The requested research product is then extracted from this initial LLM response and undergoes rule-based product review.

**Rule-based product review.** At each research step, the LLM is requested to send a response containing a specific product, with specific formatting (Fig. 1B; Tables S1,S7). Then, data-to-paper extracts the requested product from the LLM response based on its expected formatting (Tables S1,S6; for example, when requesting a "LaTex text" product, we expect the product to be enclosed within triple backticks). Failure to extract the product is translated into a feedback message sent back to the LLM (for example: "*You sent 2 triple-backtick blocks. Please send the latex as a single triple-backtick 'latex' block*"). Once the product is extracted successfully, it is programmatically refined according to a set of step-specific auto-refinement rules (Table S7, asterisk-marked rules). Then, the refined product is checked according to a set of step-specific test rules, including formatting, text length and correct referencing (for exhaustive list see Table S7). Failure to pass any of these rules is translated into a corresponding feedback message sent back to the LLM (see example in Fig. S5).

**Information tracing.** To follow information flow through all steps, data-to-paper keeps track of the specific code lines producing each file output, the translation of these outputs into tables and the incorporation of numbers from the table in the Results section. Specifically, to track numeric results in the Results section, we programmatically assign a unique label for each numeric value appearing in the prior products for the Results writing step, and present these products in the context messages with the numeric values formatted as LaTex hypertargets with their corresponding labels (Fig. 1B). We then complemented the mission prompt of the Result writing step with instructions requesting the LLM agent to wrap each numeric value that it writes with a LaTex hyperlink matching the corresponding label ("Performer mission prompt: additional instructions for data-chained manuscripts", Table S1). To allow the LLM to include numeric values which are not direct output of the code, but are rather arithmetically derived from them (like changing units, translating regression coefficients to odds ratios, etc), we further provide it with the option of using a specific syntax, \num(<formula>, "explanation"), where it can provide arithmetic formula to derive new values from values created by the code output, and provide an explanation. A rule-based feedback was added to algorithmically verify that, either as stand-alone or within a \num formula, each numeric value mentioned in the section is hyperlinked, and that the target of each link correctly matches the corresponding label provided in the prior product context. Upon compilation, the \num commands are replaced with their value and a "Notes" appendix is added listing all formulas with their explanation. To further safeguard against hallucinated or missing values, the Results "mission prompt" instructs the LLM to use a designated placeholder (specifically '[unknown]') for missing numeric values, detection of this or other placeholder in the LLM response leads to data-to-paper aborting the entire research cycle (for the list of placeholders see "Results", Table S7).

**Data-chained manuscripts.** Reflecting the tracing of information during the "Data Analysis", "Table Design" and "Results" writing steps, data-to-paper creates manuscripts that "chain" results, methods and data, where each numeric value is recursively linked to the specific lines of codes that created it. In particular, a numeric value in the "Results" section can be linked to the "Notes" appendix, and from there to a specific value in a table, and from there to the output file that was used to create this table and finally to the specific code lines which generated this output file (Supplementary Data-chained Manuscripts A-D and Supplementary Human Co-piloted Manuscripts 1-3; Note that prior manuscripts were created without this feature and do not have hyperlinks).

**Reviewer conversation.** For a subset of research steps, data-to-paper also performs an LLM review after the successful completion of rule-based product review ("Review", Fig. 1A; Fig. 2, Figs. S2,S6). LLM review is implemented in a "Reviewer conversation", which parallels the Performer conversation of the given step, but with inversion of the USER-ASSISTANT roles (Fig. 1C; Fig. 2A; see examples in Fig. 2B, Fig. S6). In parallel to its corresponding Performer conversation, this Reviewer conversation is pre-filled with the following list of context messages: (*i*) "system prompt" defining the identity of the LLM reviewer agent; (*ii*) the list of "provided prior products" for the focal step; and (*iii*) An LLM-surrogating message with the "Performer mission prompt" (namely, the "Performer mission prompt" is casted as an ASSISTANT-side message, thereby appearing as if it was created by the LLM reviewer agent). Then, the extracted product coming from the Performer conversation is presented as a USER-side message together with step-specific review instructions, in which the LLM reviewer agent is requested to choose between accepting the provided product, or providing constructive feedback ("Reviewer mission prompt", Table S1; Fig. 1C; Fig. 2A; Fig. S2). The pre-filled conversation is then sent to the LLM API to request a response from the Reviewer agent. If the Reviewer response contains feedback, it is transferred to the Performer conversation as if it were a USER-side message, requesting the LLM performer agent to provide a new response with an accordingly refined product.

**Coding steps.** For each of the three coding steps ("Data exploration", "Data analysis", "Table design"), we extract Python code (enclosed within a triple-backtick block) from the LLM response, and test this code at four levels: (*i*) *Static analysis:* Check that the code conforms to a step-specific requested structure ("Python code - Static checks", Table S7); (*ii*) *Runtime analysis:* Syntax errors, runtime errors, warnings, as well as violations of other restrictions are caught and evaluated during code execution ("Python code - Runtime checks", Table S7); (*iii*) *Package-specific guardrails:* Noting common ChatGPT coding mistakes, we wrapped the packages that we allow importing, adding multiple guardrails to monitor, control and restrict unsafe functionalities, as well as to allow rule-based review of p-value formatting (Table S8); (*iv*) *Output analysis:* Check that all the requested output files are created and contain the requested information with the requested formatting ("Numerical data checks", Table S7). Encountered issues from these 4 check levels are translated into a feedback message sent back to the LLM. As a new feedback message is added, older feedback-response message pairs are removed from the conversation (to avoid exceeding the token limits). Once the LLM-provided code passes all tests, we proceed to LLM product review: data-to-paper provides a message that shows the LLM the code output and asks it to check the code and the output and provide a list of issues and suggested corrections (see "Reviewer mission prompts" for "Data exploration" and "Data analysis" steps in Table S1). If the LLM returns suggestions for improvement, data-to-paper requests making these corrections and enters a new phase of code debugging as described above. If there are no suggestions for improvements, we end the coding step with the code and the output files it created as the corresponding research products.

**Citation retrieval.** For the two literature search steps ("Literature search I", "Literature search II", Fig. 1A; Table S1), data-to-paper augments the LLM with Semantic Scholar Academic Graph API (*28*), an external citation database and search service. This direct citation retrieval, along with algorithmic checks restricting LLM's memory-retrieved citations (Rule-based product review; Table S7), ensures that only valid citations are included in the resulting paper. These literature-search steps start with a "Devise queries" step, in which the LLM is requested to provide a list of queries for each of a pre-defined set of scopes (scopes for "Literature search I": "Dataset", "Questions"; scopes for the "Literature search II": "Background", "Dataset", "Methods", "Results"; see "Literature search I" and "Literature search II" in Table S1). Then, data-to-paper calls the citation API (*28*) to retrieve a list of citations for each of the LLM-provided queries (see example in Fig. S7). For each citation, the API provides: (*i*) *Search rank*; (*ii*) *BibTeX ID*; (*iii*) *Title*; (*iv*) *Journal and year*; (*v*) *One-sentence paper summary (TLDR) (48)*; (*vi*) *Citation influence (49)*; (*vii*) *Title and abstract embedding (40)*. Citations for each of the scopes are then filtered and sorted either by Search rank or by Title and abstract embedding similarity to the title and abstract embedding of the currently written paper (parameters in Table S9). For the runs with datasets C, D, where we attempt reproducing a specific original study, we manually excluded the citation of the original paper. The sorted lists of papers for each scope are then provided as prior products for steps in which the LLM is requested to refer to literature citations (Table S1; Fig. 1B, Fig. S7).

**LLM selection.** We compared the performance of Llama 2, Codellama and ChatGPT models in two critical research steps: (*i*) Research goal and (*ii*) Data analysis. For both tests, we used the "Health Indicators" dataset. In (*i*), we ran the research goal step of data-to-paper 10 times each either with gpt-3.5-turbo or Llama-2-70b-chat-hf, all provided with the same prior product context (Table S2). We manually annotated the goals, scoring analysis-related factors, either corresponding to true features of the dataset, or to hallucinated features not part of the dataset (Table S2, Fig. S3A). In (*ii*) we ran the data analysis step of data-to-paper 10 times each with either gpt-3.5-turbo, gpt-4, CodeLlama-34b-Instruct-hf, Llama-2-70b-chat-hf or Llama-2-7b-chat-hf and evaluated for each run the number of programmatic feedback rounds until the code passes rule-based review (Fig. S3B, Supplementary Coding Runs).

**ChatGPT models and parameters.** As the underlying LLM, we used OpenAI conversational ChatGPT models (*30*) (open-source models created hallucinated research goals and were not able to consistently converge in the data analysis coding step; LLM selection). The OpenAI models used were either gpt-3.5-turbo-0613, gpt-3.5-turbo-16k-0613, or gpt-4 (all with a knowledge cutoff of September 2021). For each research step, we assigned one of these specific models as a nominal model based on the expected conversation length of the step as well as the presumed difficulty and performance during pilot runs ("LLM", Table S1). Starting from this initial nominal model for each step, data-to-paper can automatically upgrade the model during a conversation: switching to gpt-4 when a rule-based feedback request is not resolved, and switching to gpt-3.5-turbo-16k-0613 if the number of tokens exceeds the maximum of the step's nominal model. For all models, we use default model parameters, except for the model's temperature which was specifically set for some of the steps (In particular, setting a temperature of 1 for the "Research goal" step).

**Paper assembly and compilation.** To produce the final manuscript, data-to-paper assembles a single LaTex file, combining the different manuscript-part products ("Paper assembly", Fig. 1A,B). It then automatically compiles this file, together with the list of citations retrieved from "Literature search II", into a pdf, watermarked "Created by data-to-paper (AI)".

**Manual review of created manuscripts.** We manually vetted each created manuscript and its respective run record (Supplementary Manuscripts and Runs A1-5, B1-5, C1-10, Da1-10, Db1-10, Dc1-20, Dai1-10, Dbi1-10, Dci1-10). For the manuscripts, we verified: (*i*) that the data analysis and code are correctly performed, using adequate statistical methodologies; (*ii*) that every statement in the text involving numeric information corresponds to the correct numeric value from the output of the data analysis; (*iii*) that every citation fits the context in which it was referenced; (*iv*) the overall exactness of the text; and (*v*) the quality of the overall text and wording. The manuscripts were highlighted to reflect correctly-put statements (green), imperfect, or atypical statements (yellow), minor errors (orange), and major errors (red).

**Human co-piloting.** Human co-piloting is incorporated by allowing the user to add review comments in each step after the rule-based and LLM-review have completed. If such human review is added, data-to-paper initiates a new cycle of Performer answers with rule-based checks. This process repeats iteratively until the user approves the research product of the step. We have created an app with a user interface that allows the user to follow the LLM conversation in each step and add review comments as needed.

# Data availability

The data that support the findings of this study are available in the paper and its Supplementary Information (https://github.com/rkishony/data-to-paper-supplementary).

# Code availability

Code is available at https://github.com/Technion-Kishony-lab/data-to-paper

# References

1. Z. Xi, W. Chen, X. Guo, W. He, Y. Ding, B. Hong, M. Zhang, J. Wang, S. Jin, E. Zhou, R. Zheng, X. Fan, X. Wang, L. Xiong, Y. Zhou, W. Wang, C. Jiang, Y. Zou, X. Liu, Z. Yin, S. Dou, R. Weng, W. Cheng, Q. Zhang, W. Qin, Y. Zheng, X. Qiu, X. Huang, T. Gui, The Rise and Potential of Large Language Model Based Agents: A Survey, *arXiv [cs.AI]* (2023). http://arxiv.org/abs/2309.07864.

2. Y. Liu, T. Han, S. Ma, J. Zhang, Y. Yang, J. Tian, H. He, A. Li, M. He, Z. Liu, Z. Wu, L. Zhao, D. Zhu, X. Li, N. Qiang, D. Shen, T. Liu, B. Ge, Summary of ChatGPT-Related research and perspective towards the future of large language models. *Meta-Radiology* **1**, 100017 (2023).

3. L. Wang, C. Ma, X. Feng, Z. Zhang, H. Yang, J. Zhang, Z. Chen, J. Tang, X. Chen, Y. Lin, W. X. Zhao, Z. Wei, J.-R. Wen, A Survey on Large Language Model based Autonomous Agents, *arXiv [cs.AI]* (2023). http://arxiv.org/abs/2308.11432.

4. Y. Li, D. Choi, J. Chung, N. Kushman, J. Schrittwieser, R. Leblond, T. Eccles, J. Keeling, F. Gimeno, A. Dal Lago, T. Hubert, P. Choy, C. de Masson d'Autume, I. Babuschkin, X. Chen, P.-S. Huang, J. Welbl, S. Gowal, A. Cherepanov, J. Molloy, D. J. Mankowitz, E. Sutherland Robson, P. Kohli, N. de Freitas, K. Kavukcuoglu, O. Vinyals, Competition-level code generation with AlphaCode. *Science* **378**, 1092–1097 (2022).

5. G. Spitale, N. Biller-Andorno, F. Germani, AI model GPT-3 (dis)informs us better than humans. *Sci Adv* **9**, eadh1850 (2023).

6. J. He-Yueya, G. Poesia, R. E. Wang, N. D. Goodman, Solving Math Word Problems by Combining Language Models With Symbolic Solvers, *arXiv [cs.CL]* (2023). http://arxiv.org/abs/2304.09102.

7. B. Romera-Paredes, M. Barekatain, A. Novikov, M. Balog, M. P. Kumar, E. Dupont, F. J. R. Ruiz, J. S. Ellenberg, P. Wang, O. Fawzi, P. Kohli, A. Fawzi, Mathematical discoveries from program search with large language models. *Nature* **625**, 468–475 (2024).

8. T. H. Trinh, Y. Wu, Q. V. Le, H. He, T. Luong, Solving olympiad geometry without human demonstrations. *Nature* **625**, 476–482 (2024).

9. C. Qian, X. Cong, W. Liu, C. Yang, W. Chen, Y. Su, Y. Dang, J. Li, J. Xu, D. Li, Z. Liu, M. Sun, Communicative Agents for Software Development, *arXiv [cs.SE]* (2023). http://arxiv.org/abs/2307.07924.

10. Y. Dong, X. Jiang, Z. Jin, G. Li, Self-collaboration Code Generation via ChatGPT, *arXiv [cs.SE]* (2023). http://arxiv.org/abs/2304.07590.

11. W. Zhou, Y. E. Jiang, P. Cui, T. Wang, Z. Xiao, Y. Hou, R. Cotterell, M. Sachan, RecurrentGPT: Interactive Generation of (Arbitrarily) Long Text, *arXiv [cs.CL]* (2023). http://arxiv.org/abs/2305.13304.

12. V. Nair, E. Schumacher, G. Tso, A. Kannan, DERA: Enhancing Large Language Model Completions with Dialog-Enabled Resolving Agents, *arXiv [cs.CL]* (2023). http://arxiv.org/abs/2303.17071.

13. T. Tu, A. Palepu, M. Schaekermann, K. Saab, J. Freyberg, R. Tanno, A. Wang, B. Li, M. Amin, N. Tomasev, S. Azizi, K. Singhal, Y. Cheng, L. Hou, A. Webson, K. Kulkarni, S. Sara Mahdavi, C. Semturs, J. Gottweis, J. Barral, K. Chou, G. S. Corrado, Y. Matias, A.

Karthikesalingam, V. Natarajan, Towards Conversational Diagnostic AI, *arXiv [cs.AI]* (2024). http://arxiv.org/abs/2401.05654.

14. D. A. Boiko, R. MacKnight, B. Kline, G. Gomes, Autonomous chemical research with large language models. *Nature* **624**, 570–578 (2023).

15. A. Birhane, A. Kasirzadeh, D. Leslie, S. Wachter, Science in the age of large language models. *Nature Reviews Physics* **5**, 277–280 (2023).

16. G. Conroy, How ChatGPT and other AI tools could disrupt scientific publishing. *Nature* **622**, 234–236 (2023).

17. C. Stokel-Walker, R. Van Noorden, What ChatGPT and generative AI mean for science. *Nature* **614**, 214–216 (2023).

18. C. Stokel-Walker, ChatGPT listed as author on research papers: many scientists disapprove. *Nature* **613**, 620–621 (2023).

19. M. Hutson, Could AI help you to write your next paper? *Nature* **611**, 192–193 (2022).

20. V. Berdejo-Espinola, T. Amano, AI tools can improve equity in science. *Science* **379**, 991 (2023).

21. L. Messeri, M. J. Crockett, Artificial intelligence and illusions of understanding in scientific research. *Nature* **627**, 49–58 (2024).

22. C. L. Bockting, E. A. M. van Dis, R. van Rooij, W. Zuidema, J. Bollen, Living guidelines for generative AI - why scientists must oversee its use. *Nature* **622**, 693–696 (2023).

23. E. B. Wilson, *An Introduction to Scientific Research* (Courier Corporation, 1990; https://play.google.com/store/books/details?id=rKCHDQAAQBAJ).

24. A. Teboul, Diabetes Health Indicators Dataset (2021). https://www.kaggle.com/datasets/alexteboul/diabetes-health-indicators-dataset.

25. C. G. Fink, N. Omodt, S. Zinnecker, G. Sprint, A Congressional Twitter network dataset quantifying pairwise probability of influence. *Data Brief* **50**, 109521 (2023).

26. A. L. Saint-Fleur, H. E. Alcalá, S. Sridhar, Outcomes of neonates born through meconium-stained amniotic fluid pre and post 2015 NRP guideline implementation. *PLoS One* **18**, e0289945 (2023).

27. J.-G. Shim, K.-H. Ryu, S. H. Lee, E.-A. Cho, S. Lee, J. H. Ahn, Machine learning model for predicting the optimal depth of tracheal tube insertion in pediatric patients: A retrospective cohort study. *PLoS One* **16**, e0257069 (2021).

28. R. Kinney, C. Anastasiades, R. Authur, I. Beltagy, J. Bragg, A. Buraczynski, I. Cachola, S. Candra, Y. Chandrasekhar, A. Cohan, M. Crawford, D. Downey, J. Dunkelberger, O. Etzioni, R. Evans, S. Feldman, J. Gorney, D. Graham, F. Hu, R. Huff, D. King, S. Kohlmeier, B. Kuehl, M. Langan, D. Lin, H. Liu, K. Lo, J. Lochner, K. MacMillan, T. Murray, C. Newell, S. Rao, S. Rohatgi, P. Sayre, Z. Shen, A. Singh, L. Soldaini, S. Subramanian, A. Tanaka, A. D. Wade, L. Wagner, L. L. Wang, C. Wilhelm, C. Wu, J. Yang, A. Zamarron, M. Van Zuylen, D. S. Weld, The Semantic Scholar Open Data Platform, *arXiv [cs.DL]* (2023). http://arxiv.org/abs/2301.10140.

29. Z. Ji, N. Lee, R. Frieske, T. Yu, D. Su, Y. Xu, E. Ishii, Y. Bang, W. Dai, A. Madotto, P. Fung, Survey of Hallucination in Natural Language Generation, *arXiv [cs.CL]* (2022).

  http://arxiv.org/abs/2202.03629.

30. OpenAI platform. https://platform.openai.com/docs/api-reference.

31. H. Touvron, L. Martin, K. Stone, P. Albert, A. Almahairi, Y. Babaei, N. Bashlykov, S. Batra, P. Bhargava, S. Bhosale, D. Bikel, L. Blecher, C. C. Ferrer, M. Chen, G. Cucurull, D. Esiobu, J. Fernandes, J. Fu, W. Fu, B. Fuller, C. Gao, V. Goswami, N. Goyal, A. Hartshorn, S. Hosseini, R. Hou, H. Inan, M. Kardas, V. Kerkez, M. Khabsa, I. Kloumann, A. Korenev, P. S. Koura, M.-A. Lachaux, T. Lavril, J. Lee, D. Liskovich, Y. Lu, Y. Mao, X. Martinet, T. Mihaylov, P. Mishra, I. Molybog, Y. Nie, A. Poulton, J. Reizenstein, R. Rungta, K. Saladi, A. Schelten, R. Silva, E. M. Smith, R. Subramanian, X. E. Tan, B. Tang, R. Taylor, A. Williams, J. X. Kuan, P. Xu, Z. Yan, I. Zarov, Y. Zhang, A. Fan, M. Kambadur, S. Narang, A. Rodriguez, R. Stojnic, S. Edunov, T. Scialom, Llama 2: Open Foundation and Fine-Tuned Chat Models, *arXiv [cs.CL]* (2023). http://arxiv.org/abs/2307.09288.

32. B. Rozière, J. Gehring, F. Gloeckle, S. Sootla, I. Gat, X. E. Tan, Y. Adi, J. Liu, T. Remez, J. Rapin, A. Kozhevnikov, I. Evtimov, J. Bitton, M. Bhatt, C. C. Ferrer, A. Grattafiori, W. Xiong, A. Défossez, J. Copet, F. Azhar, H. Touvron, L. Martin, N. Usunier, T. Scialom, G. Synnaeve, Code Llama: Open Foundation Models for Code, *arXiv [cs.CL]* (2023). http://arxiv.org/abs/2308.12950.

33. Q. Wu, G. Bansal, J. Zhang, Y. Wu, B. Li, E. Zhu, L. Jiang, X. Zhang, S. Zhang, J. Liu, A. H. Awadallah, R. W. White, D. Burger, C. Wang, AutoGen: Enabling Next-Gen LLM Applications via Multi-Agent Conversation, *arXiv [cs.AI]* (2023). http://arxiv.org/abs/2308.08155.

34. S. Hong, M. Zhuge, J. Chen, X. Zheng, Y. Cheng, C. Zhang, J. Wang, Z. Wang, S. K. S. Yau, Z. Lin, L. Zhou, C. Ran, L. Xiao, C. Wu, J. Schmidhuber, MetaGPT: Meta Programming for A Multi-Agent Collaborative Framework, *arXiv [cs.AI]* (2023). http://arxiv.org/abs/2308.00352.

35. A. Madaan, N. Tandon, P. Gupta, S. Hallinan, L. Gao, S. Wiegreffe, U. Alon, N. Dziri, S. Prabhumoye, Y. Yang, S. Gupta, B. P. Majumder, K. Hermann, S. Welleck, A. Yazdanbakhsh, P. Clark, Self-Refine: Iterative Refinement with Self-Feedback, *arXiv [cs.CL]* (2023). http://arxiv.org/abs/2303.17651.

36. C. Harrison, LangChain (2022). https://github.com/langchain-ai/langchain.

37. S. Gravitas, AutoGPT (2023). https://github.com/Significant-Gravitas/AutoGPT.

38. B. Peng, M. Galley, P. He, H. Cheng, Y. Xie, Y. Hu, Q. Huang, L. Liden, Z. Yu, W. Chen, J. Gao, Check Your Facts and Try Again: Improving Large Language Models with External Knowledge and Automated Feedback, *arXiv [cs.CL]* (2023). http://arxiv.org/abs/2302.12813.

39. L. Rolle-Lake, E. Robbins, "Behavioral Risk Factor Surveillance System" in *StatPearls* (StatPearls Publishing, Treasure Island (FL), 2023; https://www.ncbi.nlm.nih.gov/pubmed/31971707).

40. A. Cohan, S. Feldman, I. Beltagy, D. Downey, D. S. Weld, SPECTER: Document-level Representation Learning using Citation-informed Transformers, *arXiv [cs.CL]* (2020). http://arxiv.org/abs/2004.07180.

41. C. Steinruecken, E. Smith, D. Janz, J. Lloyd, Z. Ghahramani, "The Automatic Statistician" in *Automated Machine Learning: Methods, Systems, Challenges*, F. Hutter,

L. Kotthoff, J. Vanschoren, Eds. (Springer International Publishing, Cham, 2019; https://doi.org/10.1007/978-3-030-05318-5_9), pp. 161–173.

42. N. Altman, M. Krzywinski, P values and the search for significance. *Nat. Methods* **14**, 3–4 (2016).

43. R. Van Noorden, Hundreds of gibberish papers still lurk in the scientific literature. *Nature* **594**, 160–161 (2021).

44. G. Cabanac, C. Labbé, Prevalence of nonsensical algorithmically generated papers in the scientific literature. *J. Assoc. Inf. Sci. Technol.* **72**, 1461–1476 (2021).

45. L. Liverpool, AI intensifies fight against "paper mills" that churn out fake research. *Nature* **618**, 222–223 (2023).

46. Twitter interaction network for the US congress (2023). https://snap.stanford.edu/data/congress-twitter.html.

47. Twitter Handles for Members of the 117th Congress, *Food Research & Action Center* (2021). https://frac.org/wp-content/uploads/MOC_Twitter-Handles_117th.pdf.

48. I. Cachola, K. Lo, A. Cohan, D. S. Weld, TLDR: Extreme Summarization of Scientific Documents, *arXiv [cs.CL]* (2020). http://arxiv.org/abs/2004.15011.

49. M. Valenzuela-Escarcega, V. A. Ha, O. Etzioni, Identifying Meaningful Citations. *AAAI Workshop: Scholarly Big Data* (2015).

## Acknowledgments

We thank Ayelet Baram-Tsabari and Yael Rozenblum for discussions and providing data for initial tests, Ofer Sapir for help in organizing data-to-paper repo, Eric Lander, Yoel Fink, and Michael Elowitz for insightful discussions, and all lab members for helpful comments. LH was supported in part at the Technion by an Aly Kaufman Fellowship.

## Contributions

TI and RK conceived the study. TI and RK developed data-to-paper with inputs from LH, MK, and OA. MK and OA implemented a graphic interface for system testing. TI and LH identified and prepared the datasets and related metadata. RK and TI oversaw the autonomous research runs. TI and LH manually vetted and highlighted created papers and run files. LH conceptualized the presentation and designed the figures with inputs from TI and RK. TI, LH and RK interpreted the results and wrote the manuscript with comments from MK and OA.

## Competing interests

The authors declare no competing interests.

## Corresponding authors

Correspondence to Roy Kishony, rkishony@technion.ac.il

**Table 1.** Examples of Topics and findings of papers produced for "Health Indicators" dataset (A1-3) and "Social Network" dataset (B1-3).

| Paper | Topic and findings |
|---|---|
| A1 | **Topic:** Diabetes & physical activity<br>**Title:** "Insights into the Relationship between Physical Activity and Diabetes Prevalence"<br>**Conclusion:** "[...] a negative association between physical activity and diabetes prevalence, independent of confounders such as age, smoking status, and education level." |
| A2 | **Topic:** Physical activity & glycemic control in diabetic population<br>**Title:** "Impact of Diabetes on Physical Activity, BMI, and Demographic Factors in a Large-scale Population Study"<br>**Conclusion:** "[...] distinct patterns in physical activity levels, BMI, age distribution, and sex proportions between individuals with and without diabetes." |
| A3 | **Topic:** Diabetes & diet<br>**Title:** "The Impact of Fruit and Vegetable Consumption on Diabetes Prevalence: Insights from a Nationwide Survey"<br>**Conclusion:** "[...] significant inverse relationship between fruit and vegetable consumption and diabetes prevalence, [...]" |
| B1 | **Topic:** Congress chamber & twitter interactions<br>**Title:** "Discovering Communication Patterns in the US Congress through Twitter Interactions"<br>**Conclusion:** "[...] uncovers a significant association between the House of Representatives and the Senate regarding Twitter interactions." |
| B2 | **Topic:** Party affiliation & twitter interactions<br>**Title:** "Party Dynamics in Twitter Interactions among Members of the 117th US Congress"<br>**Conclusion:** "[...] a significant association between party affiliation and Twitter interactions, revealing higher levels of engagement within party lines." |
| B3 | **Topic:** Home state, party affiliation & twitter interaction<br>**Title:** "Insights into Social Dynamics among US Congress Members through Twitter Interactions"<br>**Conclusion:** "[...] significant distinctions in Twitter interactions among different political parties, [...] unveil the influential role of represented states." |

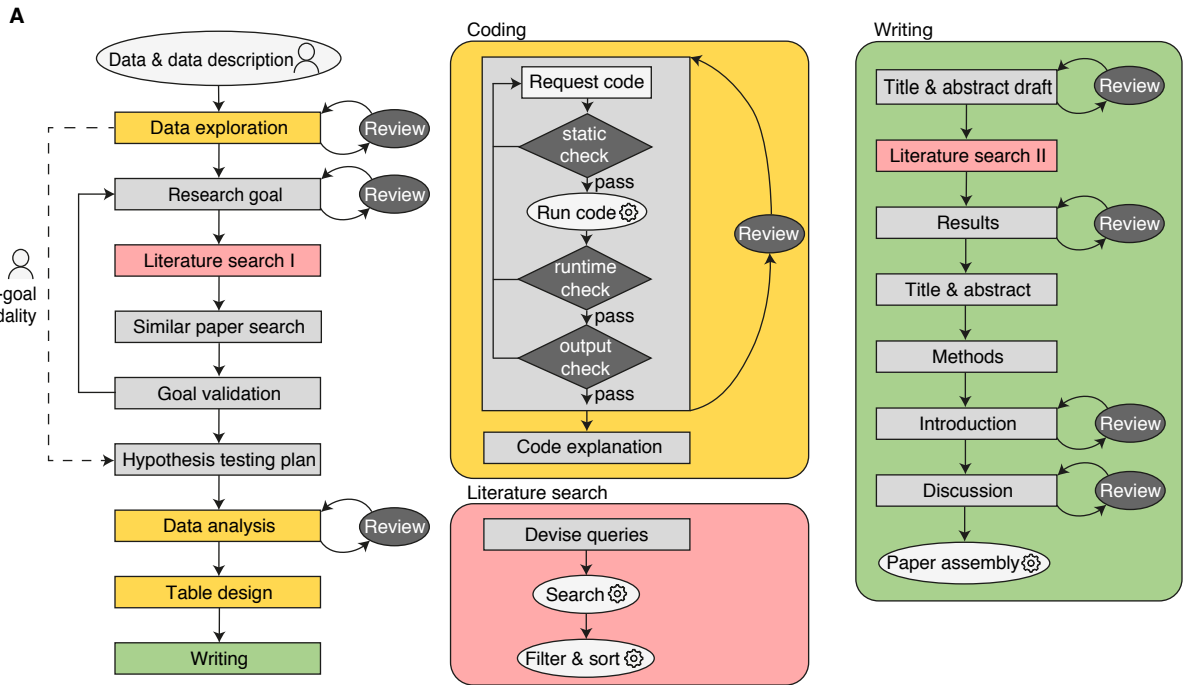
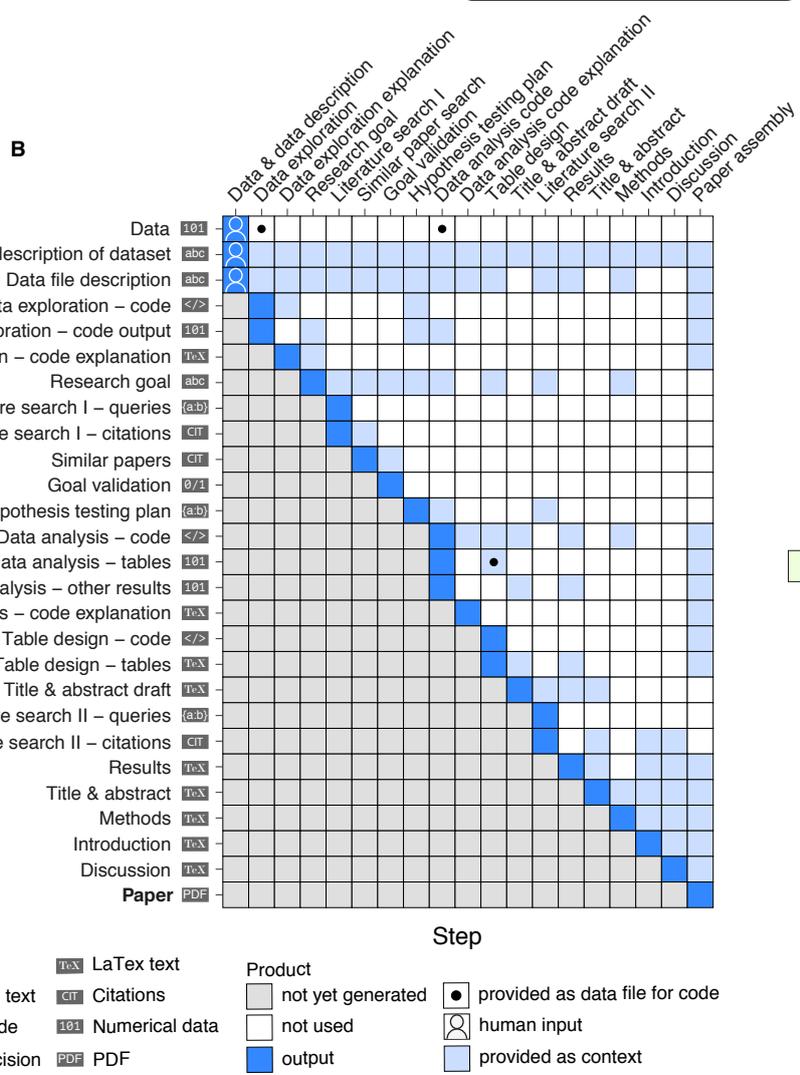
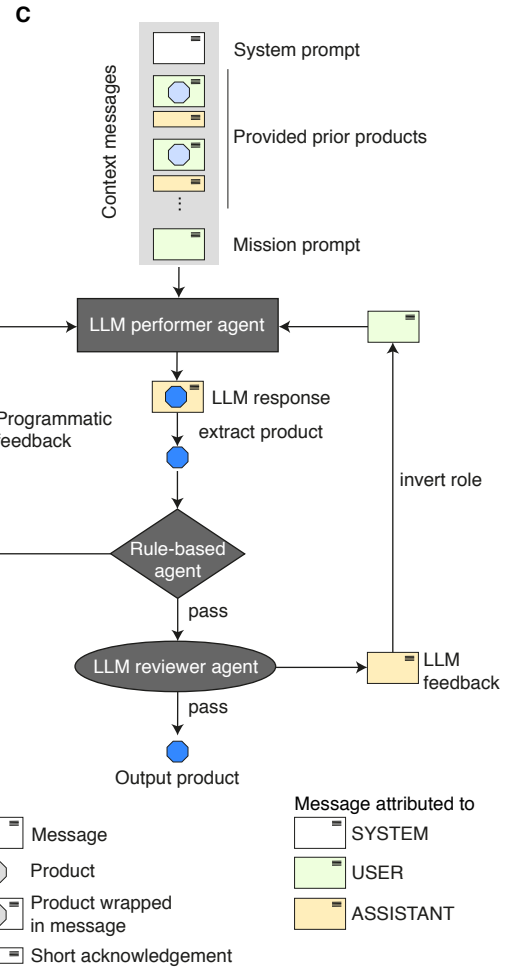

**Figure 1. data-to-paper orchestrates agent interactions and information flow through a multistep research cycle from annotated data to research paper. A.** Starting with a human-provided dataset and its textual description (ellipse with human icon), data-to-paper executes a series of LLM research steps (boxes) and programmatic tools (ellipses with gear icon), toward creating a research paper. In "fixed-goal" modality, the research goal is human provided and the goal-determining steps are skipped (dashed bypass arrow). "Coding" (yellow), "Writing" (green), and "Literature search"

(red) are modules consisting of several LLM steps and programmatic tools (Methods). Each step creates a research product that undergoes rule-based review; an example of rule-based review is shown only for the coding step, where static, runtime (including package-specific), and output checks are performed (diamond-shaped decision points; Methods). Some of the steps also incorporate an LLM review ("Review", dark ellipses; Methods). **B.** Each research step (columns) creates one or more research products (rows; "output", blue), while using a provided subset of previously created products as inputs ("provided as context", light blue). Products could be of different textual, structural, or numeric types (legend). Numeric products can be provided not only as conversation context, but also as data files for code (centered dot). The raw data files and their descriptions are provided as human input products (human icon). **C.** Each research step is implemented as a distinct LLM "Performer conversation" (Methods), programmatically filled with "context messages", including: (i) a "system prompt" message defining the agent role (white message; Table S1), (ii) "provided prior products", a series of USER-side messages (green) containing prior products (light blue octagons), each followed by a short LLM-surrogating ASSISTANT-side acknowledgment message (slim orange message), and (iii) a step-specific "mission prompt" requesting the focal product (Table S1; Figs. S1, S4; Methods). This pre-filled conversation is passed to an LLM performer agent, which generates a response from which the requested product (blue octagon) is then extracted. This product undergoes rule-based review (dark diamond) and LLM review (dark ellipse), where a response from another LLM agent is casted as if it were a USER-side response ("invert role"; Figs. S2, S6; Methods). In co-pilot mode, human review is requested after the LLM review is completed (not shown).

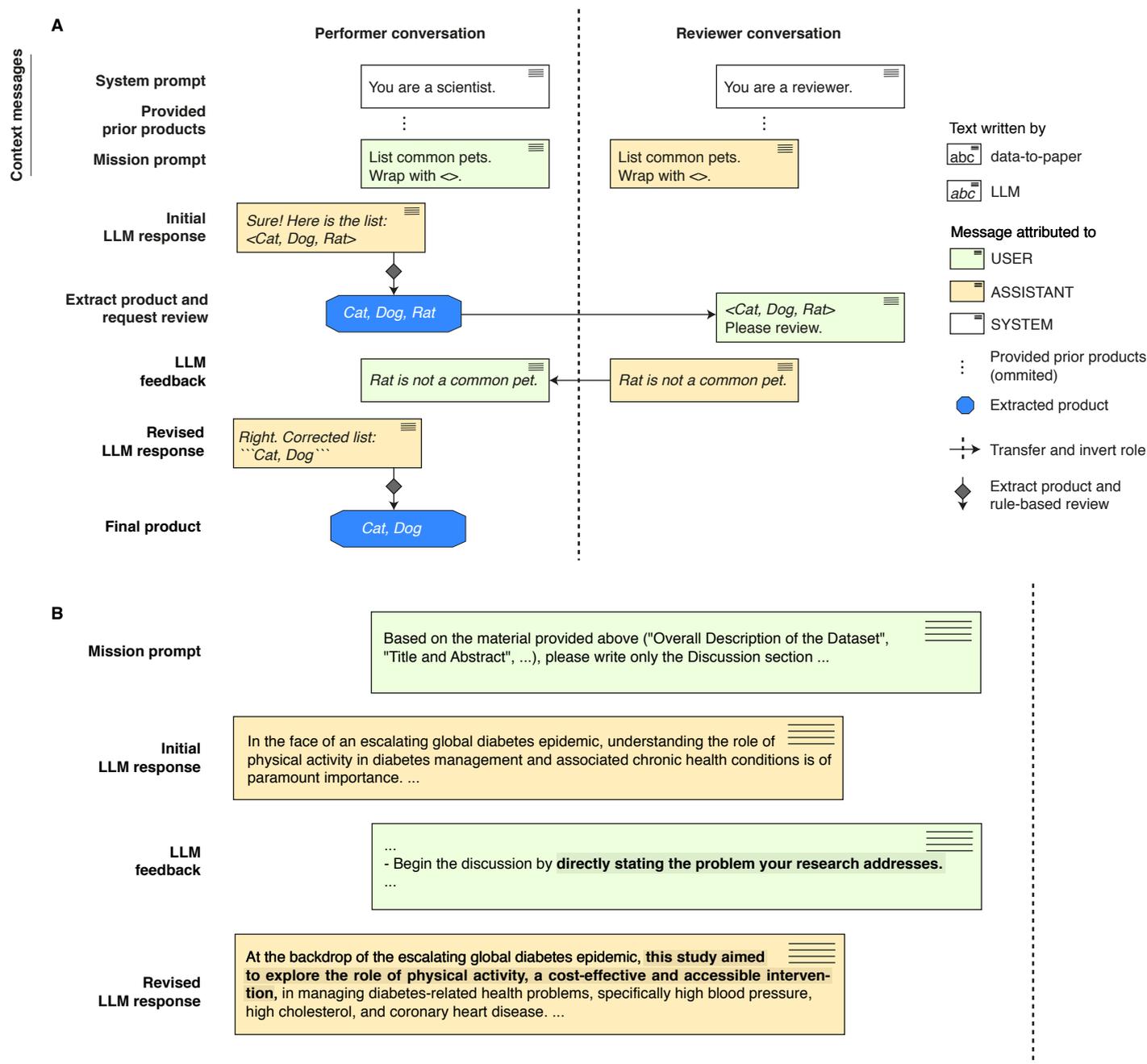

**Figure 2. Exchange among LLM agents helps refining research products. A.** Schematic illustration of LLM review. LLM review is implemented using role-inverting transfer of messages among parallel Performer (left) and Reviewer (right) conversations. Both conversations are initiated with an identity-defining "system prompt" ("System prompt", white box), followed by relevant prior products ("Provided prior products", three dots, explained in Methods, Fig. 1C, Fig. S1) and a "mission prompt" requesting the research product ("Mission prompt"). Then, an initial response containing the product is returned ("Initial LLM response"), from which the requested product is extracted and undergoes rule-based review (arrow with dark diamond; Methods). The extracted product from the ASSISTANT-side message is then transferred to the Reviewer conversation, where it is wrapped as a USER-side message ("Transfer and invert role", horizontal arrow). The LLM reviewer agent is replying with feedback ("LLM feedback", orange box) that gets role-inverted and transferred back to the Performer conversation ("LLM feedback", green box). The product is refined according to the feedback ("Revised LLM response", orange box) and gets extracted ("Final product"). **B.** Example of an interaction between performer and reviewer during Discussion writing (from Supplementary Run A5). An initial draft of a discussion paragraph ("Initial LLM response") receives reviewer comments with suggestions for improvements (bold and highlighted text, "LLM feedback"), leading to textual improvements (bold and highlighted text, "Revised LLM response").

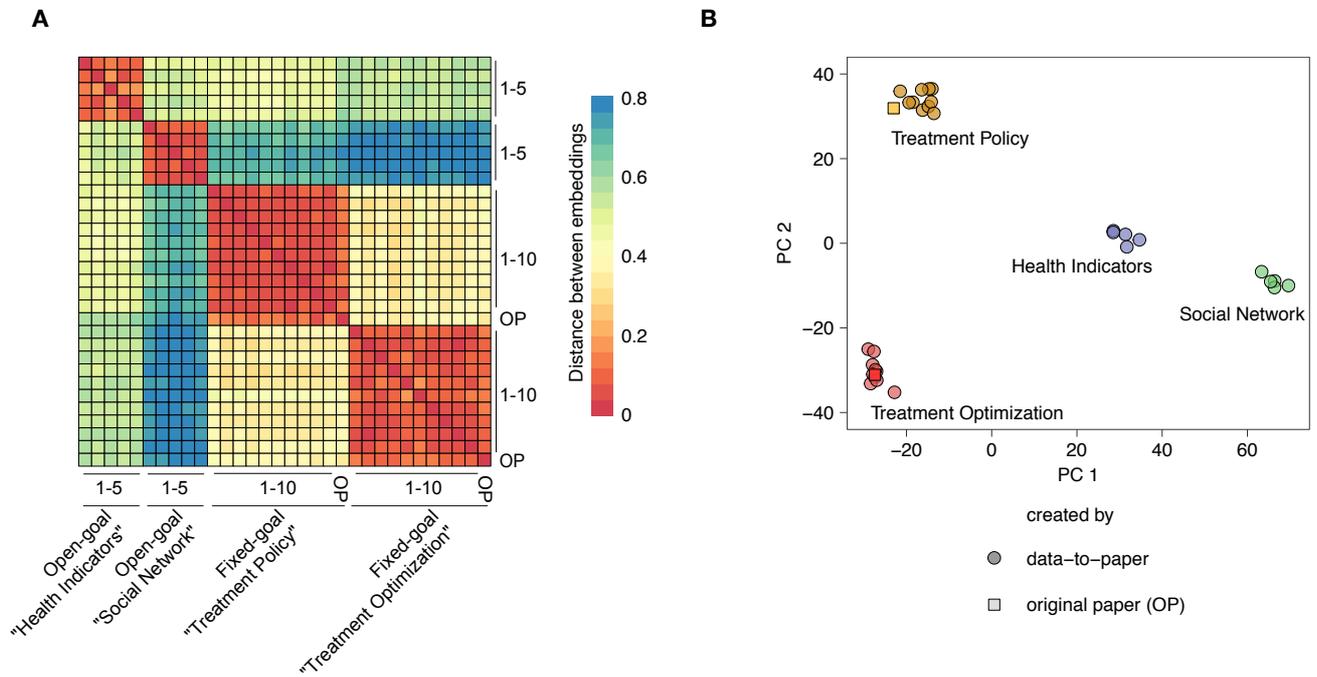

**Figure 3. Papers created from the same dataset and the respective original paper show close proximity in their vector embeddings.** Embedding vectors of the title and abstract of each paper were generated using SPECTER *(34)*. **A.** Heatmap showing pairwise cosine distance between embedding vectors of the 5 "Health Indicators" papers, 5 "Social Network" papers, 10 "Treatment Policy" papers, 10 "Treatment Optimization" papers, and 2 original papers (OP) *(20, 21)*. **B.** Principal component analysis of vector embeddings of all data-to-paper generated manuscripts (circles) for each of the 4 case studies (color; label), as well as the two original papers for the fixed-goal case studies (squares).

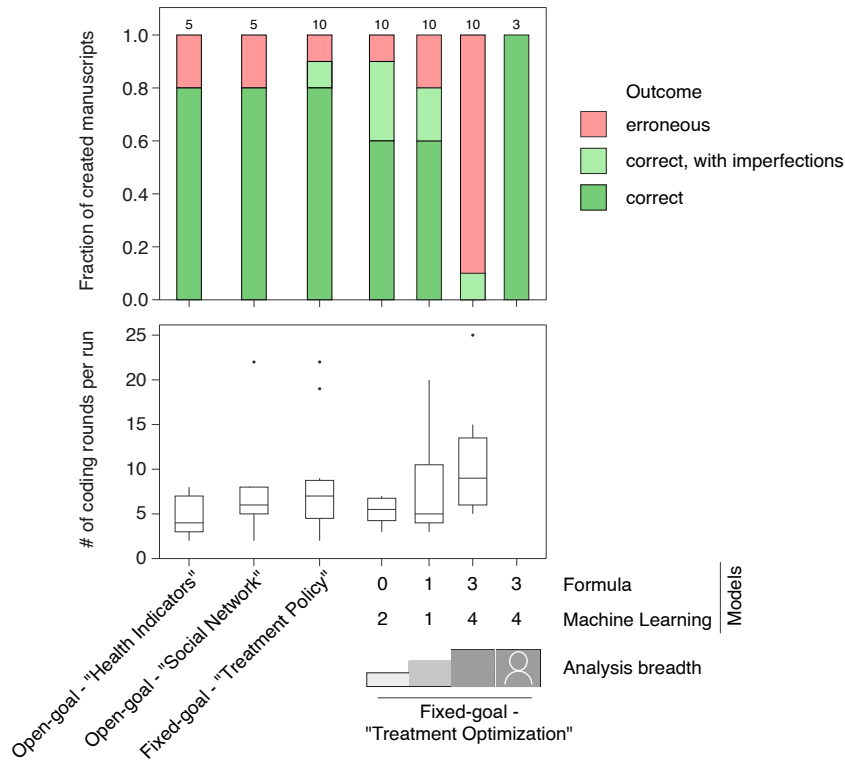

**Figure 4. Data-to-paper is able to autonomously create correct papers for simple research goals while human co-piloting is required to assure accuracy in more complex goals.** In the top, bar plot showing the fraction of correct (green), correct with imperfections (light green) and erroneous (red) papers generated for each of the 4 datasets. The numbers above the bar indicate the number of data-to-paper created papers for each set. In the bottom, the distribution of rule-based feedback iterations (see Supplementary Runs A1-5, B1-5, C1-10, Da1-10, Db1-10, Dc1-10) for each task is shown as a proxy for task complexity. For the "Treatment Optimization" dataset, 3 sets of papers were created, corresponding to research goals with varying complexity (Methods; Supplementary Data Descriptions Da, Db, Dc; Fig. S8). For the most complex goal, data-to-paper was also ran in co-pilot mode (Human icon; Supplementary Human Co-piloted Manuscripts 1-3). For the annotation of these manuscripts for imperfections and errors see text highlighting in Supplementary Manuscripts A1-5, B1-5, C1-10, Da1-10, Db1-10, Dc1-10.

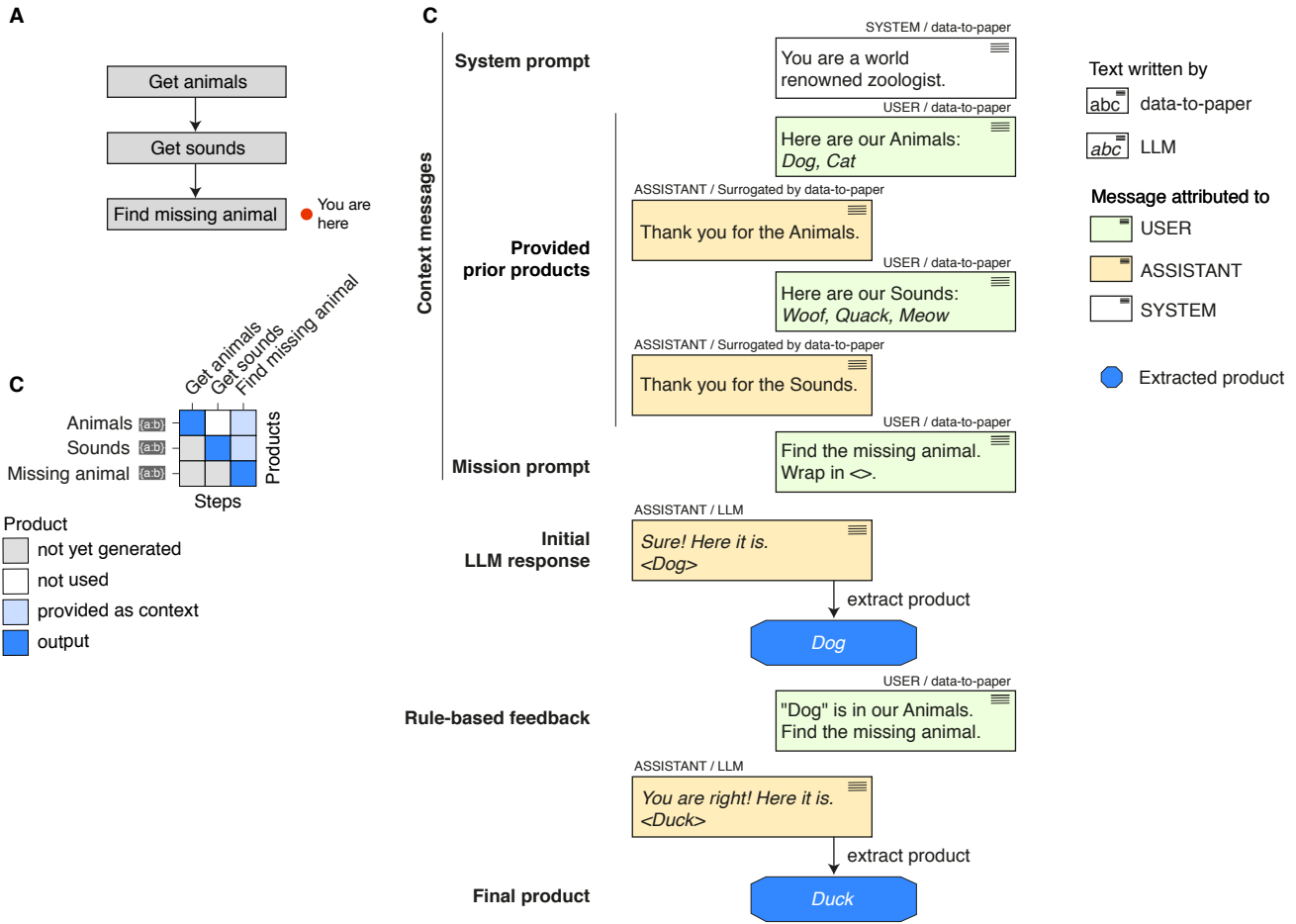

**Figure S1. Toy example of a research step implementation. A.** The toy research path consists of 3 steps: composing a list of animals, composing a list of sounds and identifying the missing animal (equivalent to Fig. 1A). **B.** Products created and used in each step. The last step of the research path is provided with the products of the two prior steps (equivalent to Fig. 1B). **C.** An implementation of the "Finding missing animal" toy research step. The conversation is filled with "context messages": (i) a "system prompt" (white message box); (ii) "provided prior products", pairs of messages where the first is a USER-side message containing one of the prior research products (green message box), and the second is a short acknowledgment message ("Thank you for the <product>", orange message box) that is programmatically written by data-to-paper (un-italicized text) yet attributed to the ASSISTANT (namely, appearing to the LLM as if it was written by it, hence LLM-surrogating; Methods); (iii) "mission prompt" indicating the requested product (here, name of the missing animal) and the formatting (here enclosed with <>; see real formatting instructions in Table S1). This programmatically filled conversation is sent to the LLM API, which returns an LLM-authored response message (italicized text, orange message box). The research product is then automatically extracted from this LLM response according to its predefined formatting (blue octagon). The extracted product undergoes rule-based checks, and upon failure an appropriate feedback message is sent to the LLM as a USER-side message ("Rule-based feedback", green message box). The LLM replies with a corrected response, from which the final product is extracted (blue octagon). See also a real example of a research step in Fig. S4.

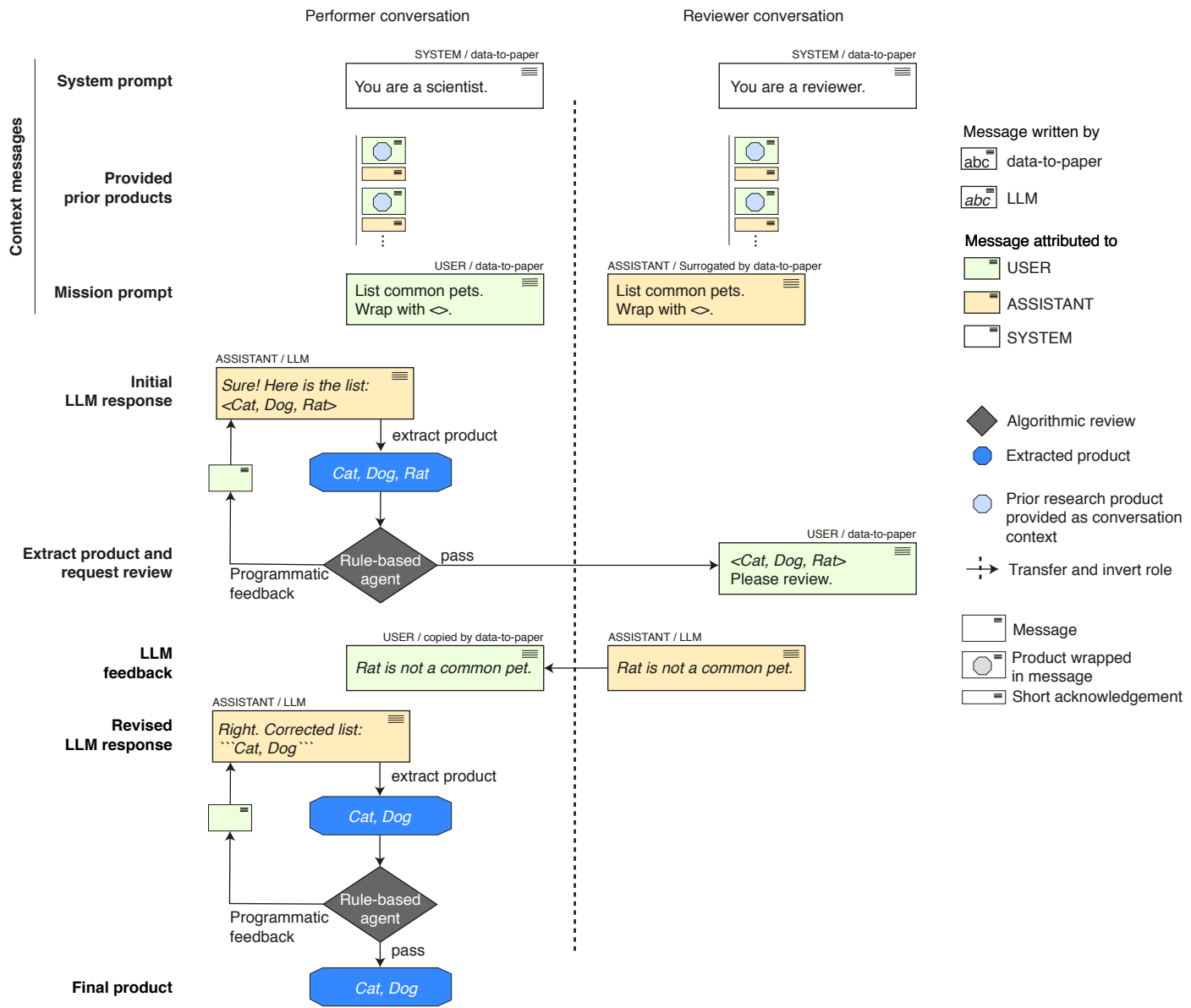

**Figure S2. Internal LLM review is implemented by transferring messages between two role-inverted LLM conversations.** Layout of the Performer conversation (left) and the parallel role-inverted Reviewer conversation (right; see also Fig. 2A, Methods). First, the two conversations are programmatically filled with a list of "context messages", including a "system prompt", messages providing prior products, and the "mission prompt" (as in Fig. 1C, Fig. S1; Methods). Notably, the mission prompt is added as a USER-side message (green message box) in the Performer conversation and as an LLM-surrogating ASSISTANT message (orange message box) in the Reviewer conversation (Methods). Second, data-to-paper requests an LLM response for the Performer conversation ("Initial LLM response", orange box), extracts the requested product, and performs rule-based checks, providing programmatic feedback (upwards going green message box; Methods; Fig. S1). Third, once the product passes rule-based review, it is sent for LLM review; a USER-side message containing the extracted product and review instructions is appended to the reviewer conversation ("Extract product and request review", green message box; Supplementary Table 1), and a response is requested from the LLM reviewer ("LLM feedback", orange message box). Fourth, data-to-paper copies the review message, appending it to the performer conversation as if it were a USER-side message ("LLM feedback", green message box). Finally, a new LLM response is requested from the performer, in which it corrects the product according to the feedback from the LLM reviewer agent ("Revised LLM response", orange message box). The revised product undergoes rule-based review, until it passes, which terminates the step with a final product.

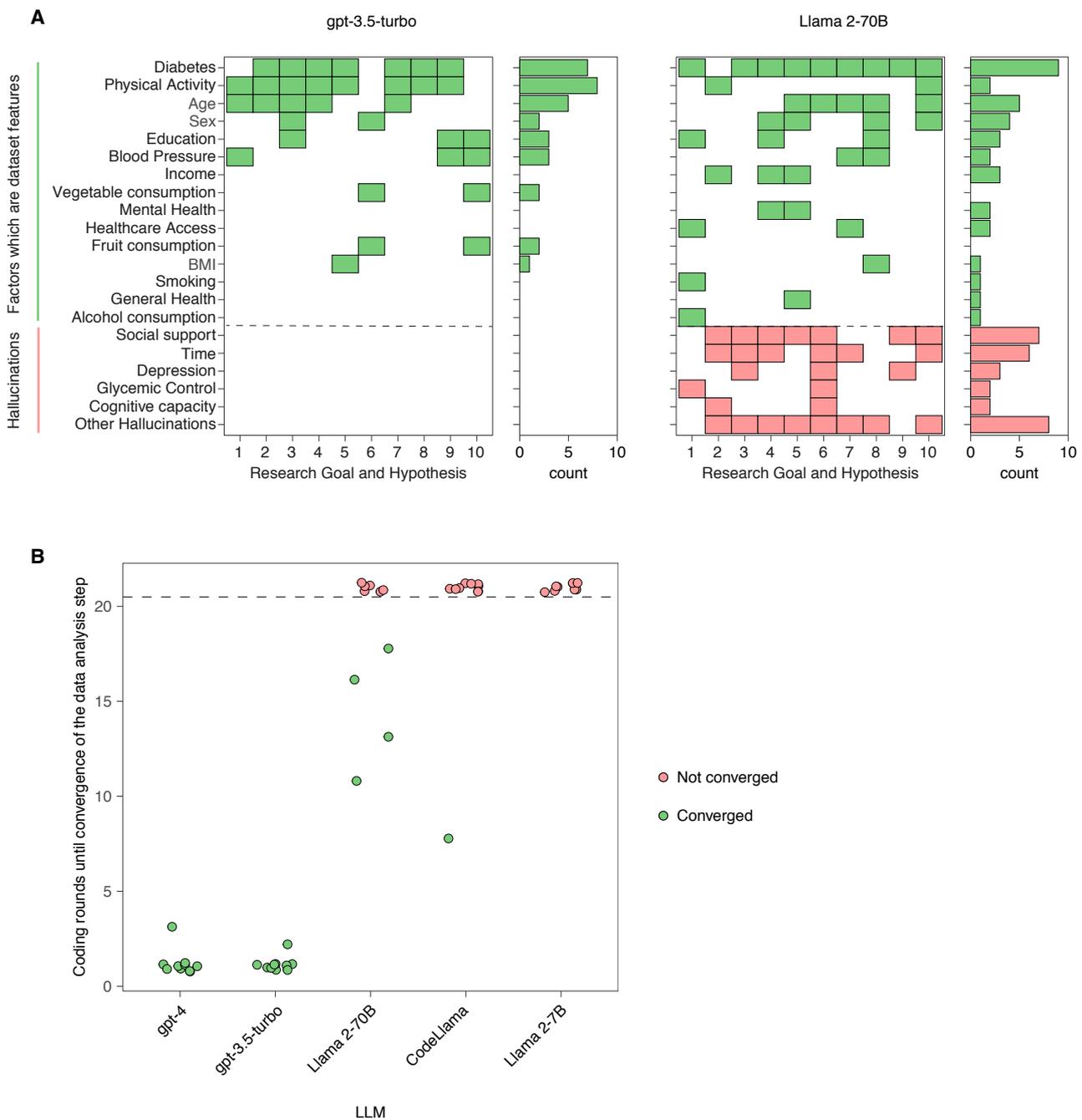

**Figure S3. Current state-of-the-art open-source LLMs are not able to perform key data-to-paper steps. A.** Evaluation of the performance of different LLMs in the "Research goal" step. Starting from the same "context messages", including Data exploration results, we ran the "Research goal" step on the "Health Indicators" dataset for 10 times with either gpt-3.5-turbo or Llama 2-70B, and manually vetted the resulting research goals, annotating factors as either being dataset features or hallucinations (Supplementary Table 2). While gpt-3.5-turbo only used factors present in the dataset in the research goal and hypotheses, Llama 2-70B hallucinated in all 10 research, providing goals that included factors or information which were not present in the dataset. **B.** Evaluation of LLMs in the "Data analysis" step. Each LLM performed the "Data analysis" step on the "Health Indicators" dataset for 10 times with programmatic review only. A maximum of 20 coding rounds was allowed. While gpt-3.5-turbo and gpt-4 converged to functional code within 1-3 coding rounds (median = 1), Llama 2-7B never converged, CodeLlama converged only once after 8 coding rounds, and Llama 2-70B in only 4 runs with 13-18 coding rounds (Supplementary Coding Runs).

**Conversation start**

**Context messages**

**System prompt**

SYSTEM / data-to-paper

You are a scientist who needs to write literature search queries.

**Provided prior producs**

USER / data-to-paper — Message wrapping research product

Here are the title and abstract of the paper:

\title{Impact of Diabetes on Physical Activity, BMI, and Demographic Factors in a Large-scale Population Study}

\begin{abstract}
Diabetes is a prevalent chronic health condition with significant implications for public health. Understanding the impact of diabetes on physical activity, body mass index (BMI), and demographic factors is crucial for effective interventions. However, ...

— Product provided as conversation context / First element of the Conversation context

ASSISTANT / Surrogated by data-to-paper

Thank you for the Title and the Abstract.

**Other provided products** ⋮

**Mission prompt**

USER / data-to-paper

Please write literature-search queries that we can use to search for papers related to our study. — General instructions

You would need to compose search queries to identify prior papers covering these 4 areas:

"background": papers that provide background on the overall subject of our study
"dataset": papers that use the same or similar datasets as in our study
"methods": papers that use the same or similar methods as in our study
"results": papers that report results similar to our study

Return your answer as a `Dict[str, List[str]]`, where the keys are the 4 areas noted above, and the values are lists of query string. Each individual query should be a string with up to 5-10 words.

— Detailed instructions on output research product

For example, for a study reporting waning of the efficacy of the covid-19 BNT162b2 vaccine based on analysis of the "United Kingdom National Core Data (UK-NCD)", the queries could be:
{
"background": ['SARS-CoV2 spread', 'covid-19 global impact', 'covid-19 vaccine'],
"dataset": ['The UK-NCD dataset', 'covid-19 vaccine efficacy dataset'],
"methods": ['covid-19 vaccine efficacy analysis', 'kaplan-meier survival analysis'],
"results": ['covid-19 vaccine efficacy', 'covid-19 vaccine efficacy over time', 'covid-19 vaccine waning']
}

— One-shot example of the output

**LLM response**

ASSISTANT / ChatGPT

Here are the literature search queries for each area:

{
    "background": ['Diabetes prevalence worldwide', 'Impact of diabetes on physical health', 'Effects of physical activity on glycemic control', 'Relationship between BMI and diabetes'],
    "dataset": ['Behavioral Risk Factor Surveillance System (BRFSS)', 'CDC BRFSS data analysis', 'Large-scale population study on diabetes and physical activity'],
    "methods": ['Multiple linear regression analysis on glycemic control', 'Covariate adjustment in linear regression', 'Associations between physical activity and demographic factors in diabetes'],
    "results": ['Association between physical activity and glycemic control in diabetes', 'Effect of age and sex on physical activity levels in diabetes', 'Impact of BMI on glycemic control in diabetes']
}

— Output research product wrapped in LLM response

**Research product**

{
    "background": ['Diabetes prevalence worldwide', 'Impact of diabetes on physical health', 'Effects of physical activity on glycemic control', 'Relationship between BMI and diabetes'],
    "dataset": ['Behavioral Risk Factor Surveillance System (BRFSS)', 'CDC BRFSS data analysis', 'Large-scale population study on diabetes and physical activity'],
    "methods": ['Multiple linear regression analysis on glycemic control', 'Covariate adjustment in linear regression', 'Associations between physical activity and demographic factors in diabetes'],
    "results": ['Association between physical activity and glycemic control in diabetes', 'Effect of age and sex on physical activity levels in diabetes', 'Impact of BMI on glycemic control in diabetes']
}

— data-to-paper extracts the conversation product (here the literature search queries)

**Figure S4. An example of a research step conversation.** An example Performer conversation for the "Literature search II" step for the "Health Indicators" dataset (Supplementary Run A2), composed of the following list of messages (letterheaded boxes, titled with "attribution / source"; attributions: ASSISTANT, USER or SYSTEM; sources: data-to-paper or ChatGPT): "context messages", including (i) "system prompt", a SYSTEM message defining the identity of the LLM agent; (ii) "provided prior products", a list of USER-ASSISTANT message pairs providing relevant products of prior steps (here, for simplicity, only the "Title & abstract" prior product is shown out of the 4 context products provided in this step, Fig. 1B). To mimic a conversation, each USER-side message providing a product is followed by an LLM-surrogating acknowledgment message; (iii) "mission prompt", a USER-side message providing instructions for the LLM, including general instructions, detailed instructions and formatting instructions (Table S1). This programmatically filled conversation is sent to OpenAI ChatGPT API (24). Based on the prior product context and the specific instructions for the current step, the LLM then generates a response message that should contain the requested research product (here a dictionary containing the list of queries for each requested scope). The LLM-created research product is automatically extracted from the message by data-to-paper (blue octagonal box). These extracted products then further undergo rule-based and LLM-based review (Fig. 1C; Fig. 2A; Figs. S1, S2, S5, S6).

**A**

*Mission prompt* — USER / data-to-paper

Please write literature-search queries that we can use to search for papers related to our study.
...
Return your answer as a `Dict[str, List[str]]`, where the keys are the 4 areas noted above, and the values are lists of query string. Each individual query should be **a string with up to 5-10 words.**

*Initial LLM response* — ASSISTANT / ChatGPT

```
{...
"results": ['Association between physical activity and glycemic control in diabetes', 'Effect of age and sex on physical activity levels in diabetes', 'Impact of BMI on glycemic control in diabetes']
...
}
```

*Rule-based feedback* — USER / data-to-paper

Queries should be 5-10 word long.

The following queries are too long:
**"Effect of age and sex on physical activity levels in diabetes"**

Please return your complete response again, with these queries shortened.

*Revised LLM response* — ASSISTANT / ChatGPT

Apologies for the confusion. Here is the revised version with shorter queries:
```
{...
 "results": ['physical activity and glycemic control', 'age and sex impact on physical activity', 'BMI and glycemic control in diabetes'
...
}
```

**B**

*Mission prompt* — USER / data-to-paper

Write a complete Python code to analyze the data and create dataframes as basis for scientific Tables for our paper.

The code must have the following sections (with these exact capitalized headers):

`# IMPORT`
...
**# SAVE ADDITIONAL RESULTS**
...

*Initial LLM response* — ASSISTANT / ChatGPT

```python
# IMPORT
...
additional_results = {
 'Total number of observations': len(df_cleaned),
 'Pseudo R-squ of regression model': result.prsquared
}
with open('additional_results.pkl', 'wb') as f:
    pickle.dump(additional_results, f)
...
```

*Rule-based feedback* — USER / data-to-paper

Your code must contain the following sections:
('# IMPORT', '# LOAD DATA', '# DATASET PREPARATIONS', '# DESCRIPTIVE STATISTICS', '# PREPROCESSING', '# ANALYSIS', '# SAVE ADDITIONAL RESULTS').
But I could not find these headers:
['# SAVE ADDITIONAL RESULTS'].
Please rewrite the complete code again with all the required sections.

*Revised LLM response* — ASSISTANT / ChatGPT

I apologize for the oversight. Here is the corrected code:

```python
# IMPORT
...
# SAVE ADDITIONAL RESULTS
additional_results = {
 'Total number of observations': len(df_cleaned),
 'Pseudo R-squ of regression model': result.prsquared
}
with open('additional_results.pkl', 'wb') as f:
    pickle.dump(additional_results, f)
...
```

**Figure S5. Example of rule-based review.** Two examples are shown for algorithmic rule-based review of products extracted from LLM responses (Supplementary Manuscripts A2,4). **A.** Devising literature-search queries. In the "Literature search II" step, a "mission prompt" message requests the LLM to write search queries, for each of several scopes, indicating the allowed length of these queries ("Mission prompt"; Table S1). The LLM performer agent returns a message containing a list of queries for each scope ("Initial LLM response"). Detecting that one of these queries is violating the length limitation, the rule-based agent issues an algorithmic feedback message to be sent back to the LLM, requesting to shorten this specific query ("Rule-based feedback", bold and highlighted text). The LLM is then replying with a set of new queries, now properly adhering to the length limitation (bold and highlighted text, "Revised LLM response"). **B.** Data analysis code. A "mission prompt" requests to write code for data analysis, specifying the format of the code and its output (bold and highlighted text, "Mission prompt"). LLM responds with a code that omits one of the required headers ("Initial LLM response"). Static code check by data-to-paper catches that mistake and relays feedback response to the LLM, requesting to rewrite the code with all required sections, specifically mentioning the missing header (bold and highlighted text, "Rule-based feedback"). The LLM then resends the corrected code with the additional required header (bold and highlighted text, "Revised response").

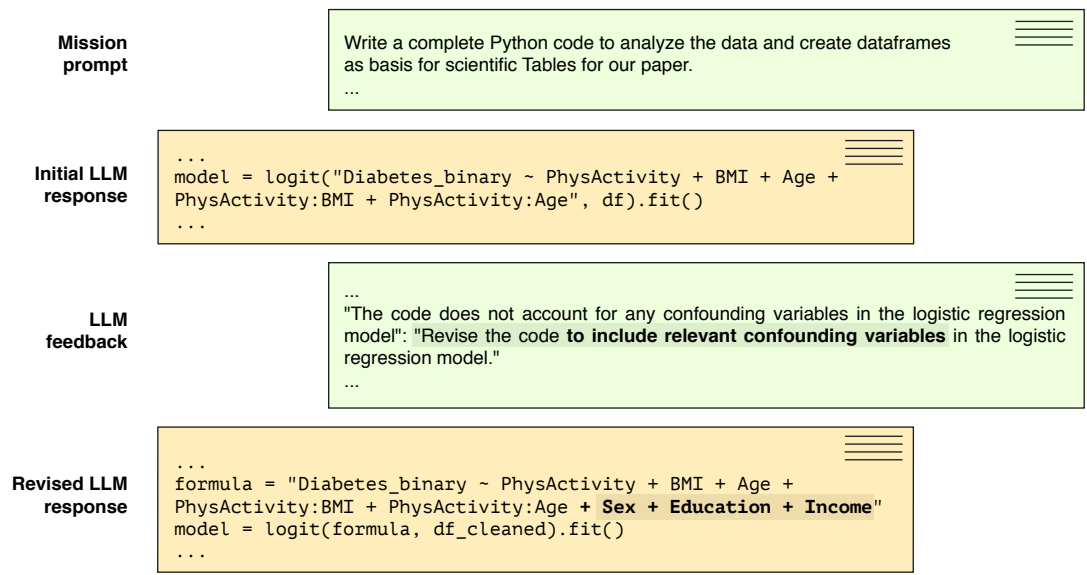

**Figure S6. Example of LLM review.** An initial code (Supplementary Manuscript A4), created in the "Data analysis" step, tests for interactions between variables, but does not adequately account for confounding factors ("Initial LLM response"). Following an LLM review feedback indicating this problem (bold and highlighted text, "LLM feedback"), the model is adequately augmented with additional confounding factors (bold and highlighted text, "Revised response").

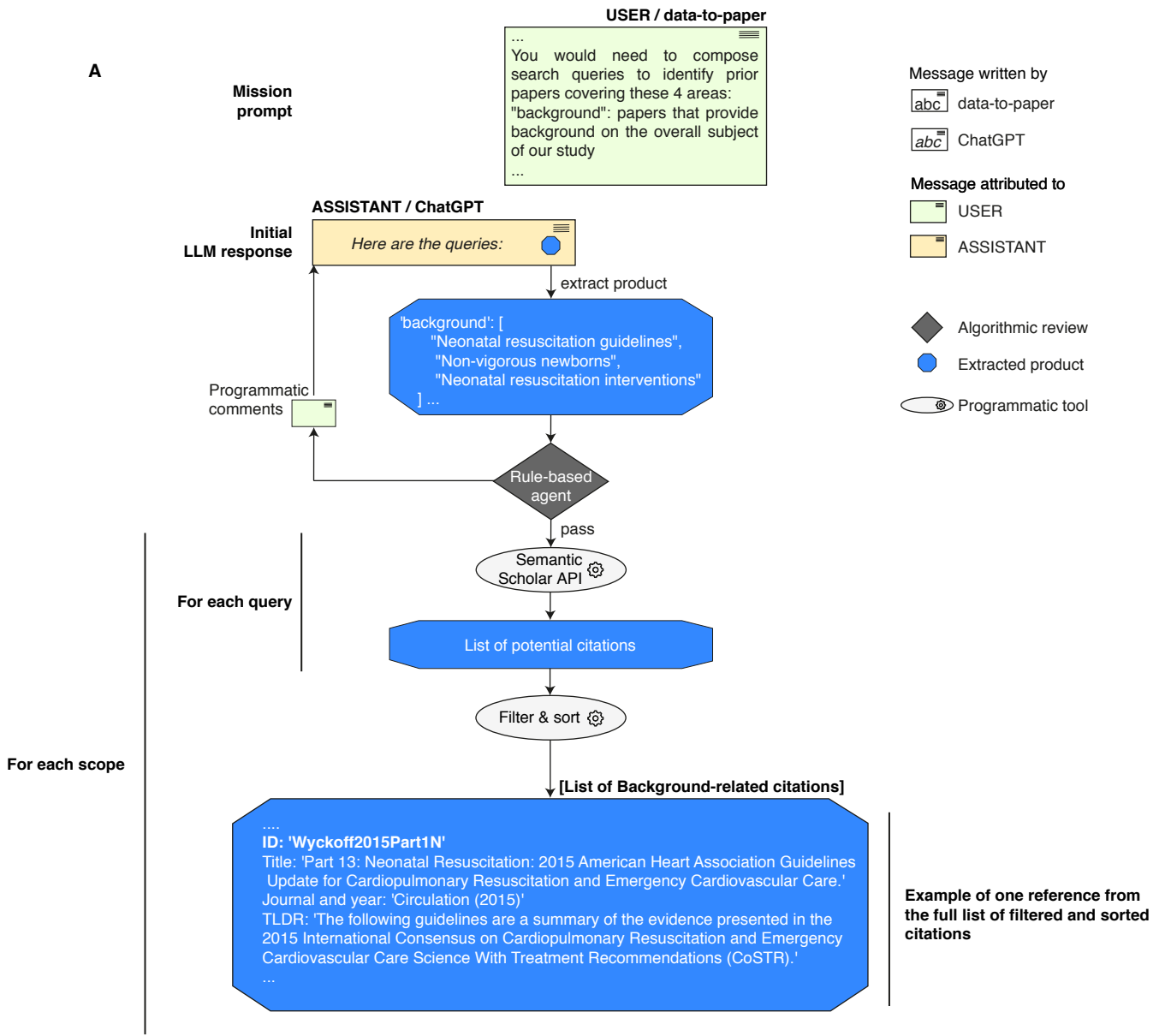

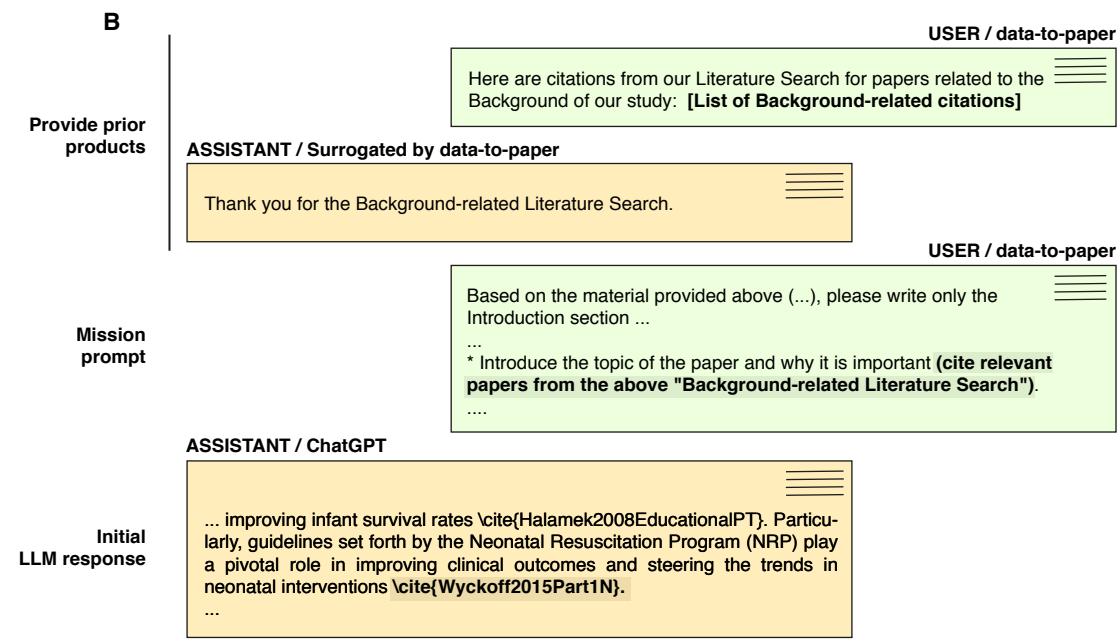

**Figure S7. An external citation search API is used to retrieve citations based on LLM-created search queries.** Instead of using the LLM to directly retrieve citations from memory (which can lead to hallucinated citations), we use it to devise literature search queries, which data-to-paper uses to perform a programmatic literature search using an external API (22). **A.** "Mission prompt" requests literature search queries related to the "Title & abstract draft" on four different areas: "Background", "Dataset", "Results" and "Methods" in a "Dict[str, List[str]]" format ("Mission prompt", green box). The LLM performer agent provides these queries in an accordingly-formatted message ("Initial LLM response", orange box), from which they are extracted by data-to-paper (blue octagon) and passed to a rule-based review (dark diamond; Table S7). Once the queries pass rule-based review, each query is used to retrieve papers through a call to the Semantic Scholar Academic Graph API (22) (light ellipse with a gear icon). The returned queries are then programmatically filtered and sorted (light ellipse with gear icon; Methods; Table S9), resulting in the final research product (blue octagon). **B.** An example of the "Introduction" writing step, showing how previously obtained citation lists are provided as part of prior products (bold text, "Provided prior products", green box). The performer agent is then requested to cite relevant papers from the provided list (bold and highlighted text, "Mission prompt"), which it does by using the appropriate citation BibTeX ID in the text (bold and highlighted text, bold and highlighted text, "Initial LLM response").

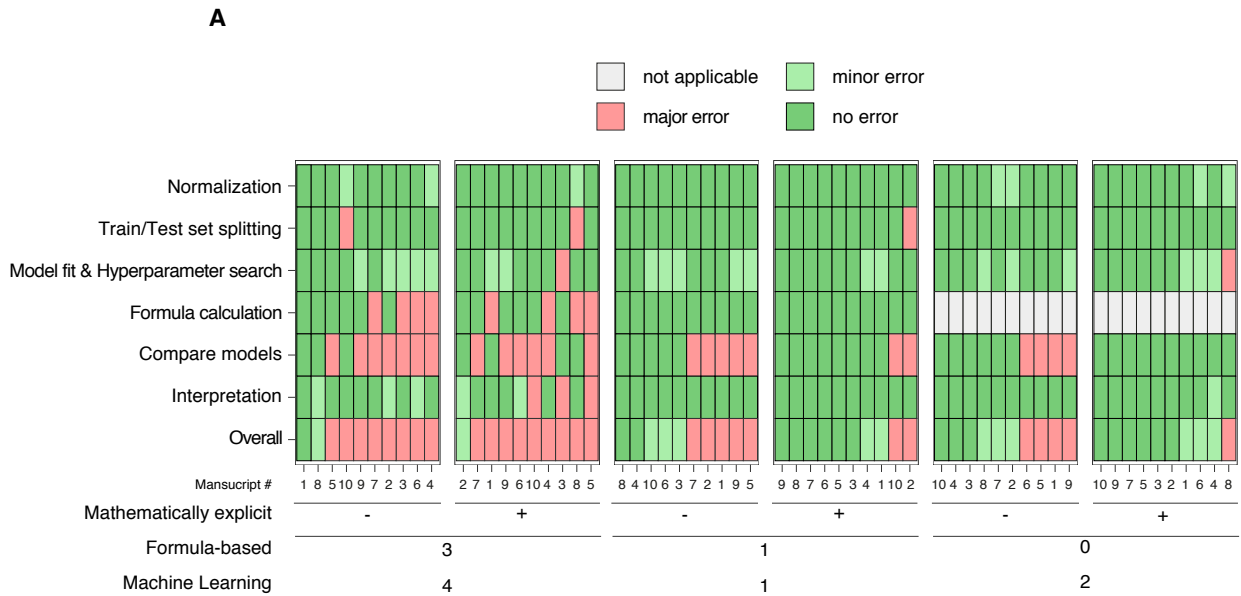

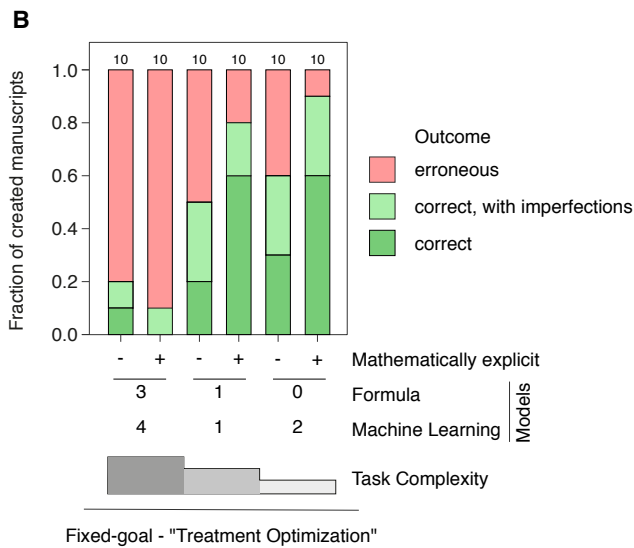

**Figure S8. Increasing goal complexity leads to erroneous analysis by data-to-paper. A.** Evaluation of different analysis and interpretation tasks (rows) for all data-to-paper "Treatment Optimization" runs (columns), for different research goals, varying in the number of Machine Learning and Formula models they require to create, as well as in whether they provide mathematically explicit explanations (see corresponding research goals in Supplementary Data Descriptions Dai, Da, Dbi, Db, Dci, Dc). Tasks are scored as correct (green), correct with imperfections (light green) and erroneous papers (red; corresponding to text highlighting in Supplementary Manuscripts Dai1-10, Da1-10, Dbi1-10, Db1-10, Dci1-10, Dc1-10). Within each goal, the 10 created manuscripts are ordered according to their number of erroneous tasks (original manuscript numbers are indicated). The "Overall" row indicates an error in any of the analysis or interpretation tasks. **B.** An overall summary of the fraction of correct, correct with imperfections and erroneous papers for each of the 6 goals.